\documentclass[aip,jcp,reprint]{revtex4-1}
\pdfoutput=1
\usepackage{amsmath}
\usepackage{braket}
\usepackage{graphicx}
\usepackage{hyperref}
\usepackage{color}

\begin{document}
\title{\color{blue}{DMRG-CASPT2 study of the longitudinal static second hyperpolarizability of all-trans polyenes}}

\author{Sebastian Wouters}
\affiliation{Center for Molecular Modelling, Ghent University, Technologiepark 903, 9052 Zwijnaarde, Belgium}

\author{Veronique {Van Speybroeck}}
\affiliation{Center for Molecular Modelling, Ghent University, Technologiepark 903, 9052 Zwijnaarde, Belgium}

\author{Dimitri {Van Neck}}
\affiliation{Center for Molecular Modelling, Ghent University, Technologiepark 903, 9052 Zwijnaarde, Belgium}

\begin{abstract}
We have implemented internally contracted complete active space second order perturbation theory (CASPT2) with the density matrix renormalization group (DMRG) as active space solver [\href{http://dx.doi.org/10.1063/1.3629454}{Y. Kurashige and T. Yanai, \textit{J. Chem. Phys.} \textbf{135}, 094104 (2011)}].
Internally contracted CASPT2 requires to contract the generalized Fock matrix with the 4-particle reduced density matrix (4-RDM) of the reference wavefunction.
The required 4-RDM elements can be obtained from 3-particle reduced density matrices (3-RDM) of different wavefunctions, formed by symmetry-conserving single-particle excitations op top of the reference wavefunction.
In our spin-adapted DMRG code \textsc{chemps2} [\url{https://github.com/sebwouters/chemps2}], we decompose these excited wavefunctions as spin-adapted matrix product states, and calculate their 3-RDM in order to obtain the required contraction of the generalized Fock matrix with the 4-RDM of the reference wavefunction.
In this work, we study the longitudinal static second hyperpolarizability of all-trans polyenes C$_{2n}$H$_{2n+2}$ [$n=4-12$] in the cc-pVDZ basis set. DMRG-SCF and DMRG-CASPT2 yield substantially lower values and scaling with system size compared to RHF and MP2, respectively.
\end{abstract}

\maketitle

\section{Introduction}

Non-linear optical (NLO) properties of organic conjugated materials have been studied extensively due to their potential use in optical devices.
In computational chemistry, it is well known that approximate methods struggle to yield qualitatively accurate NLO properties.
Density functional theory (DFT) with the commonly used exchange-correlation functionals significantly overestimates the longitudinal static second hyperpolarizability $\gamma_{zzzz}$ and its scaling with system size.\cite{JCP.109.10489, PhysRevLett.83.694, JPCA.104.4755}
Long-range corrected functionals reduce a large part of the overestimation, indicating the importance of long-range exchange for the NLO properties.\cite{JCP.119.11001, JCP.122.234111, JCP.126.014107, JCP.129.024117}

Also in \textit{ab initio} methods, electron correlation should be included with care.\cite{PhysRevA.20.1313}
Restricted Hartree-Fock (RHF) theory overestimates $\gamma_{zzzz}$, and with second order M{\o}ller-Plesset perturbation theory (MP2) this overestimation becomes even worse.\cite{CPL.457.276}
For short polyenes, coupled-cluster theory with single and double excitations (CCSD) predicts larger $\gamma_{zzzz}$ values compared to RHF, but their ratio drops below 1 with increasing system length.\cite{JCP.135.014111}
The effect of electron correlation on $\gamma_{zzzz}$ has since been further explored with various methods.\cite{JCP.136.024315, JCP.137.054301}

One particular method, the density matrix renormalization group (DMRG),\cite{PhysRevLett.69.2863} is especially well-suited to study one-dimensional systems such as hydrogen chains and polyenes.
It has therefore been used to assess the accuracy of other methods to obtain $\gamma_{zzzz}$.\cite{CPL.457.276, JCP.136.134110}

White invented DMRG in 1992 to solve the breakdown of Wilson's numerical renormalization group for real-space lattice systems.\cite{PhysRevLett.69.2863}
A few years later, \"Ostlund and Rommer discovered its underlying wavefunction ansatz, the matrix product state (MPS).\cite{PRL.75.3537}
White and Martin applied DMRG for the first time to \textit{ab initio} Hamiltonians in 1999.\cite{JCP.110.4127}
With the effort of several quantum chemistry groups,\cite{JCP.115.6815, JCP.116.4462, PRB.67.125114, JCP.122.024107, JCP.128.014107, JCP.130.234114, PRB.81.235129, CPC.185.1501, JCP.143.244118} DMRG has since become a standard method in computational chemistry.

For realistic systems and Hamiltonians, applying DMRG to the entire system becomes infeasible.
DMRG can then be used as a numerically exact active space solver in conjunction with the complete active space self consistent field method (DMRG-SCF) to treat the static correlation.\cite{JCP.128.144116, JCP.128.144117, IJQC.109.2178, JCP.140.241103}
Dynamic correlation can be added subsequently by canonical transformation theory,\cite{JCP.132.024105, PCCP.14.7809} internally contracted perturbation theory,\cite{JCP.135.094104, JCP.141.174111, JCTC.12.1583} or a configuration interaction expansion.\cite{JCP.139.044118, JCTC.11.5120}
Alternatively, the perturbation wavefunctions can also be solved within the DMRG framework.\cite{JCP.141.111101, JCP.143.102815, JCP.144.034103}
Recently DMRG has shown its full potential by resolving the electronic structure of important catalysts, containing multiple transition metal atoms.\cite{NC.5.660, NC.6.927, JACS.136.15977}

The MPS ansatz and the DMRG algorithm are reviewed in Sect.~\ref{section_mps_ansatz} and \ref{section_dmrg_method}, respectively.
Our implementation of the contraction of the generalized Fock matrix with the 4-particle reduced density matrix (4-RDM) of the reference wavefunction is described in Sect.~\ref{our_implementation}.
The DMRG-SCF algorithm and internally contracted complete active space second order perturbation theory (CASPT2) are outlined in Sect.~\ref{section_dmrg_active_space_solver}.
In Sect.~\ref{gamma_for_polyenes} we use our implementation of DMRG-CASPT2 to study the longitudinal static second hyperpolarizability of all-trans polyenes, and compare the results with lower levels of theory.
A summary of this work is given in Sect.~\ref{section_conclusions}.

\section{MPS ansatz} \label{section_mps_ansatz}
Starting from a full configuration interaction (FCI) wavefunction for $L$ spatial orbitals
\begin{multline}
\ket{\Psi} = \sum\limits_{\{ n_{i \sigma} \}} C^{ n_{1 \uparrow} n_{1\downarrow} n_{2\uparrow} ... n_{L\downarrow} } \\  \left( \hat{a}^{\dagger}_{1\uparrow} \right)^{n_{1\uparrow}} \left( \hat{a}^{\dagger}_{1\downarrow} \right)^{n_{1\downarrow}}  \left( \hat{a}^{\dagger}_{2\uparrow} \right)^{n_{2\uparrow}}  ... \left( \hat{a}^{\dagger}_{L\downarrow} \right)^{n_{L\downarrow}}  \ket{ - },
\end{multline}
the MPS ansatz can be introduced. Without loss of accuracy, the FCI coefficient tensor can be decomposed as
\begin{eqnarray}
& & C^{ n_{1 \uparrow} n_{1\downarrow} n_{2 \uparrow} n_{2\downarrow} n_{3 \uparrow} n_{3\downarrow} ... n_{L \uparrow} n_{L\downarrow} } \nonumber \\
& = & \sum\limits_{\{ \alpha_i \}} A[1]^{ n_{1\uparrow} n_{1\downarrow} }_{\alpha_0, \alpha_1} A[2]^{ n_{2\uparrow} n_{2\downarrow} }_{\alpha_1,\alpha_2} A[3]^{ n_{3\uparrow} n_{3\downarrow} }_{\alpha_2,\alpha_3} ... A[L]^{ n_{L\uparrow} n_{L\downarrow} }_{\alpha_{L-1}, \alpha_L} \nonumber \\
& = & \mathbf{A}[1]^{ n_{1} } \mathbf{A}[2]^{ n_{2} } \mathbf{A}[3]^{ n_{3} } ... \mathbf{A}[L]^{ n_{L} },
\end{eqnarray}
with $\text{size}(\alpha_i) = \min( 4^{i}, 4^{L-i})$ to allow to represent every state of the Hilbert space.
On the last line the shorthand $n_i$ for $n_{i\uparrow}n_{i\downarrow}$ was introduced. Matrices are written in boldface.
The MPS ansatz can now be obtained by truncating $\text{size}(\alpha_i)$ to $\min( 4^{i}, 4^{L-i}, D )$.
The parameter $D$ is called the \textit{virtual} or \textit{bond} dimension of the MPS.
With increasing virtual dimension $D$, a larger corner of the Hilbert space can be reached, and this parameter is therefore a handle on the convergence.

Because of the virtual dimension truncation, the MPS ansatz is not invariant with respect to orbital rotations.
With arguments from quantum information theory, it can be shown that for one-dimensional orbital spaces, any accuracy per length unit can be reached with a finite virtual dimension $D$, independent of the number of orbitals $L$, even if $L \rightarrow \infty$.\cite{JSMTE.2007.P08024} 
In quantum chemistry, the orbital active spaces are often far from one-dimensional.
With a proper choice and ordering of the active space orbitals, and by exploiting the symmetry group of the Hamiltonian, efficient MPS decompositions of the FCI coefficient tensor can still be obtained.\cite{EPJD.68.272}

The symmetry group of the Hamiltonian typically consists of $\mathsf{SU(2)}$ spin symmetry, $\mathsf{U(1)}$ particle number symmetry, and the spatial point group symmetry $\mathsf{P}$.
In \textsc{chemps2} all three Hamiltonian symmetries are exploited, but $\mathsf{P}$ is limited to the abelian point groups with real-valued character tables.\cite{CPC.185.1501}
The non-abelian $\mathsf{SU(2)}$ spin symmetry was first exploited in condensed matter and nuclear structure DMRG calculations,\cite{NPB.495.505, EPL.57.852, PRC.73.014301, PRB.78.245109} and later found its way to quantum chemistry.\cite{JCP.128.014107, JCP.130.234114, JCP.136.124121, JCP.136.134110, CPC.185.1501, JCP.144.134101}
We also want to mention Refs.~\onlinecite{JCP.140.104112} and \onlinecite{JCP.142.024107}, where the non-abelian point group $\mathsf{D_{\infty h}}$ is exploited for homonuclear diatomic molecules.

In order to exploit the symmetry, both the orbital and virtual indices are written with the correct symmetry labels.
In what follows $j$ and $s$ denote spin, $j^z$ and $s^z$ denote spin projection, $N$ denotes particle number, and $I$ denotes an irreducible representation of an abelian point group $\mathsf{P}$ with real-valued character table.
In that case $I \otimes I = I^{\text{trivial}}$ for all $I$.
Due to the Wigner-Eckart theorem, the MPS tensors factorize into reduced tensors and Clebsch-Gordan coefficients:
\begin{multline}
A[i]^{s s^z N I}_{ j_L j_L^z N_L I_L \alpha_L ; j_R j_R^z N_R I_R \alpha_R} = \braket{ j_L j_L^z s s^z \mid j_R j_R^z } \\
 \delta_{N_L + N, N_R} \delta_{I_L \otimes I, I_R } T[i]^{s N I}_{ j_L N_L I_L \alpha_L; j_R N_R I_R \alpha_R }. \label{symmMPS}
\end{multline}
The Clebsch-Gordan coefficients introduce block sparsity and information compression.
A reduced basis state $(j_L N_L I_L \alpha_L)$ in the reduced MPS tensor $\mathbf{T}[i]^{sNI}$ represents an entire multiplet $(j_L j_L^z N_L I_L \alpha_L)$ in the MPS tensor $\mathbf{A}[i]^{ss^zNI}$.
The reduced virtual dimension is therefore smaller than the virtual dimension it represents:
\begin{multline}
 D^{\text{reduced}} = \sum\limits_{ j_L N_L I_L }D^{\text{reduced}}_{ j_L N_L I_L } \\ 
 < D^{\text{represented}} = \sum\limits_{ j_L N_L I_L } ( 2 j_L + 1 ) D^{\text{reduced}}_{ j_L N_L I_L }.
\end{multline}
Another advantage of the symmetry-adaptation is the ability to study excited states as ground states of different symmetry sectors, for example to calculate singlet-triplet gaps.
We also want to mention that the decomposition in Eq.~\eqref{symmMPS} resembles the wavefunction in the multifacet graphically contracted function method.\cite{JCP.141.064105}

When a non-singular matrix $\mathbf{G}$ and its inverse are inserted at a particular virtual bond, the MPS tensors change but the wavefunction does not:
\begin{eqnarray}
\mathbf{A}[i]^{n_i} \mathbf{A}[i+1]^{n_{i+1}} & = & \left( \mathbf{A}[i]^{n_i} \mathbf{G} \right) \left( \mathbf{G}^{-1} \mathbf{A}[i+1]^{n_{i+1}} \right) \nonumber \\
 & = & \widetilde{\mathbf{A}}[i]^{n_i} \widetilde{\mathbf{A}}[i+1]^{n_{i+1}}.
\end{eqnarray}
This gauge freedom allows to left-normalize
\begin{equation}
\sum\limits_{ n_{i} } \left( \mathbf{A}[i]^{n_{i}} \right)^{\dagger} \mathbf{A}[i]^{n_{i}} = \mathbf{I}
\end{equation}
or right-normalize
\begin{equation}
\sum\limits_{ n_{i} } \mathbf{A}[i]^{n_{i}} \left( \mathbf{A}[i]^{n_{i}} \right)^{\dagger} = \mathbf{I}
\end{equation}
the MPS tensors.
Analogous expressions exist for reduced MPS tensors.\cite{CPC.185.1501} Left- and right-normalization are used in the DMRG algorithm to remove the overlap matrix from the local generalized eigenvalue problems.

\section{DMRG algorithm} \label{section_dmrg_method}

First, a so-called \textit{micro-iteration} or local eigenvalue problem of the DMRG algorithm is discussed. Subsequently, its place in the total DMRG algorithm is explained. A micro-iteration corresponds to the optimization of two neighbouring MPS tensors, while all other MPS tensors remain fixed. Suppose the MPS tensors belong to orbitals $i$ and $i+1$. A two-orbital tensor
\begin{equation}
\mathbf{B}[i]^{n_i n_{i+1}} = \mathbf{A}[i]^{n_i} \mathbf{A}[i+1]^{n_{i+1}}
\end{equation}
is constructed by contracting over the common virtual index.
It forms an initial guess for the local eigenvalue problem which arises by varying the Lagrangian
\begin{equation}
\mathcal{L} = \braket{ \Psi( \mathbf{B}[i] ) \mid \hat{H} \mid \Psi( \mathbf{B}[i] ) } - E \braket{ \Psi( \mathbf{B}[i] ) \mid \Psi( \mathbf{B}[i] ) }
\end{equation}
with respect to the parameters in $\mathbf{B}[i]^{n_i n_{i+1}}$. When all MPS tensors to the left of orbital $i$ are left-normalized and all MPS tensors to the right of orbital $i+1$ are right-normalized, this variation yields a standard eigenvalue problem:
\begin{equation}
\sum\limits_{\widetilde{n}_i \widetilde{n}_{i+1}} \mathbf{H}^{n_i n_{i+1}}_{\widetilde{n}_i \widetilde{n}_{i+1}} \times \mathbf{B}[i]^{\widetilde{n}_i \widetilde{n}_{i+1}} = E \mathbf{B}[i]^{n_i n_{i+1}}.\label{effective_ham_equation}
\end{equation}
In practice, the application of the sparse effective Hamiltonian on the left-hand side of Eq.~\eqref{effective_ham_equation} is realized by:
\begin{multline}
\sum\limits_{\widetilde{n}_i \widetilde{n}_{i+1}} \mathbf{H}^{n_i n_{i+1}}_{\widetilde{n}_i \widetilde{n}_{i+1}} \times \mathbf{B}[i]^{\widetilde{n}_i \widetilde{n}_{i+1}} \\ = \sum_k \sum\limits_{\widetilde{n}_i \widetilde{n}_{i+1}} \mathbf{O}[i,k]^{n_i }_{\widetilde{n}_i} \mathbf{B}[i]^{\widetilde{n}_i \widetilde{n}_{i+1}} \mathbf{C}[i,k]^{n_{i+1} }_{\widetilde{n}_{i+1}},
\end{multline}
where the sum over $k$ runs over $\mathcal{O}(L^2)$ operator and complementary operator pairs.
The lowest eigenvalue and corresponding $\mathbf{B}[i]^{n_i n_{i+1}}$ tensor of the effective Hamiltonian are typically obtained with the Lanczos or Davidson algorithm. This tensor is then decomposed with a singular value decomposition into a left-normalized MPS tensor $\mathbf{A}[i]^{n_i}$, a right-normalized MPS tensor $\mathbf{A}[i+1]^{n_{i+1}}$, and $4\times\min( \text{size}(\alpha_{i-1} ),\text{size}(\alpha_{i+1} ))$ Schmidt values $\mathbf{\Lambda}$:
\begin{equation}
\mathbf{B}[i]^{n_{i} n_{i+1}} = \mathbf{A}[i]^{n_{i}} \mathbf{\Lambda} \mathbf{A}[i+1]^{n_{i+1}}.
\end{equation}
When the number of Schmidt values is larger than $D$, only the $D$ largest ones are kept.

This local eigenvalue equation is solved during so-called left and right \textit{sweeps} or \textit{macro-iterations}.
During a left (right) sweep, the $D$ largest Schmidt values are absorbed into the MPS tensor $\mathbf{A}[i]^{n_i}$ ($\mathbf{A}[i+1]^{n_{i+1}}$). In the next step, the MPS tensors of orbitals $i-1$ and $i$ ($i+1$ and $i+2$) are then optimized.
The DMRG sweeps continue until the energy and/or the wavefunction does not change anymore or a maximum number of sweeps is reached.
The virtual dimension truncation $D$ is stepwise increased after several sweeps.

Analogous expressions exist for the reduced two-orbital tensor, its corresponding effective Hamiltonian equation, and the subsequent decomposition into left- and right-normalized reduced MPS tensors.\cite{CPC.185.1501}
For more details on the DMRG algorithm we refer the reader to Refs.~\onlinecite{JCP.116.4462, EPJD.68.272}.
Per sweep, its computational cost is $\mathcal{O}(L^3D^3 + L^4D^2)$, its memory requirement $\mathcal{O}(L^2D^2)$, and its disk requirement $\mathcal{O}(L^3D^2)$.
In what follows, $D$ will always denote the reduced virtual dimension.

The operators
\begin{eqnarray}
 \hat{b}^{\dagger}_{i \sigma} & = & \hat{a}^{\dagger}_{i \sigma} \label{operator_dagger}\\
 \hat{b}_{i \sigma} & = & \left( -1 \right)^{\frac{1}{2} - \sigma} \hat{a}_{i~-\sigma} \label{operator_nodagger}
\end{eqnarray}
are doublet irreducible tensor operators, belonging to respectively row $(s = \frac{1}{2}, s^z = \sigma, N=1, I_i)$ of irreducible representation $(s = \frac{1}{2}, N=1, I_i)$ and row $(s = \frac{1}{2}, s^z = \sigma, N=-1, I_i)$ of irreducible representation $(s = \frac{1}{2}, N=-1, I_i)$. The Wigner-Eckart theorem can therefore be exploited for both the MPS tensors and the second quantized operators. The resulting $\mathsf{SU(2)}$ Clebsch-Gordan coefficients can be removed by summing over the common multiplets, and only Wigner 6-j and 9-j symbols remain in a spin-adapted DMRG code.\cite{CPC.185.1501}

\begin{figure*}[t]
 \includegraphics[width=0.93\textwidth]{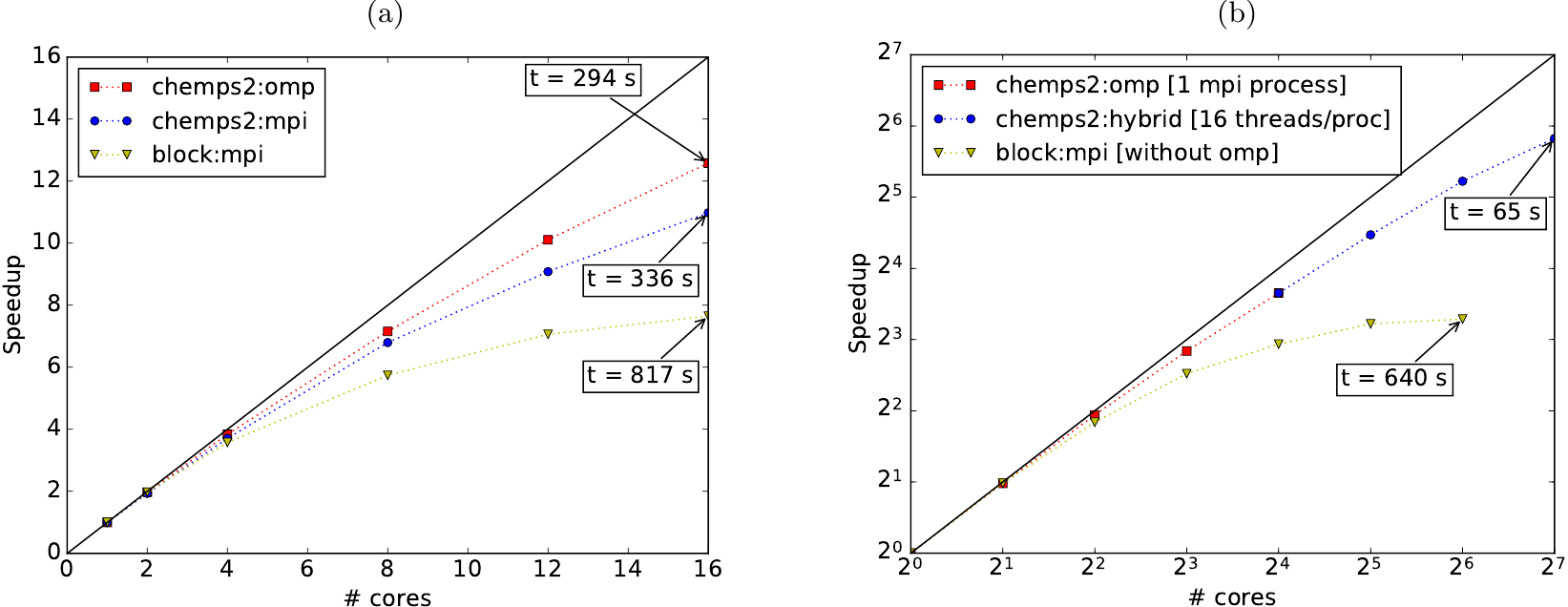}
 \caption{\label{speedup_figure} Speedup achieved with the hybrid parallelization of \textsc{chemps2}, compared with the MPI parallelization of \textsc{block} version 1.1-alpha.\cite{block-1.1-alpha} The system under study is H$_2$O in Roos' ANO DZ basis set ($L = 41$) with atom positions O(0,0,0) and H($\pm$0.790689766, 0, 0.612217330) in \AA.\cite{JCP.118.8551} All calculations were performed with reduced virtual dimension $D=1000$. RHF orbitals were used, ordered per irreducible representation of $\mathsf{C_{2v}}$ according to their single-particle energy, and the irreducible representations were ordered as $\{ \mathsf{A1},~\mathsf{A2},~\mathsf{B1},\text{~and~}\mathsf{B2} \}$. The residual norm tolerance for the Davidson algorithm was set to $10^{-4}$. Note that in \textsc{block} the square of this parameter needs to be passed. Each node has a dual Intel Xeon Sandy Bridge E5-2670 (total of 16 cores at 2.6 GHz) and 64 GB of memory. The nodes are connected with FDR InfiniBand. The renormalized operators were stored on GPFS in order to achieve high disk bandwidths. Both codes and all depending libraries were compiled with the Intel MPI compiler version 2015.1.133. The Intel Math Kernel Library version 11.2.1.133 was used for BLAS and LAPACK routines. (a) Comparison of pure MPI and OMP speedups on a single node. Wall times per sweep are indicated for 16 cores (in seconds). (b) Illustration of the hybrid parallelization of \textsc{chemps2}. For 16 cores and less, one MPI process with several OMP threads is used. For 32 cores and more, several MPI processes each with 16 OMP threads are used. Wall times per sweep are indicated (in seconds).}
\end{figure*}

We have reviewed the symmetry-adaption and the use of operator and complementary operator pairs in order to explain the hybrid parallelization in \textsc{chemps2} for mixed distributed and shared memory architectures. MPI processes\cite{mpi_reference} become responsible for certain operator and complementary operator pairs. This strategy was first introduced in Ref.~\onlinecite{JCP.120.3172}. The contraction over separate reduced symmetry sectors $(j_L N_L I_L)$ is parallelized in \textsc{chemps2} with OMP threads.\cite{CPC.185.1501, omp_reference} The parallelization over symmetry sectors was implemented in Ref.~\onlinecite{JCP.130.234114} for distributed memory architectures. The parallelization over operator and complementary operator pairs is independent from the parallelization over symmetry sectors, and they give independent (multiplicative) speedups. Our hybrid parallelization is illustrated in Fig.~\ref{speedup_figure}. The caption of Fig.~\ref{speedup_figure} contains all details of the calculation. We also want to mention a third parallelization strategy for DMRG, which involves multiple simultaneous sweeps.\cite{PRB.87.155137}

\section{Contraction of the generalized Fock matrix and the 4-RDM} \label{our_implementation}
\begin{figure}[t]
 \includegraphics[width=0.46\textwidth]{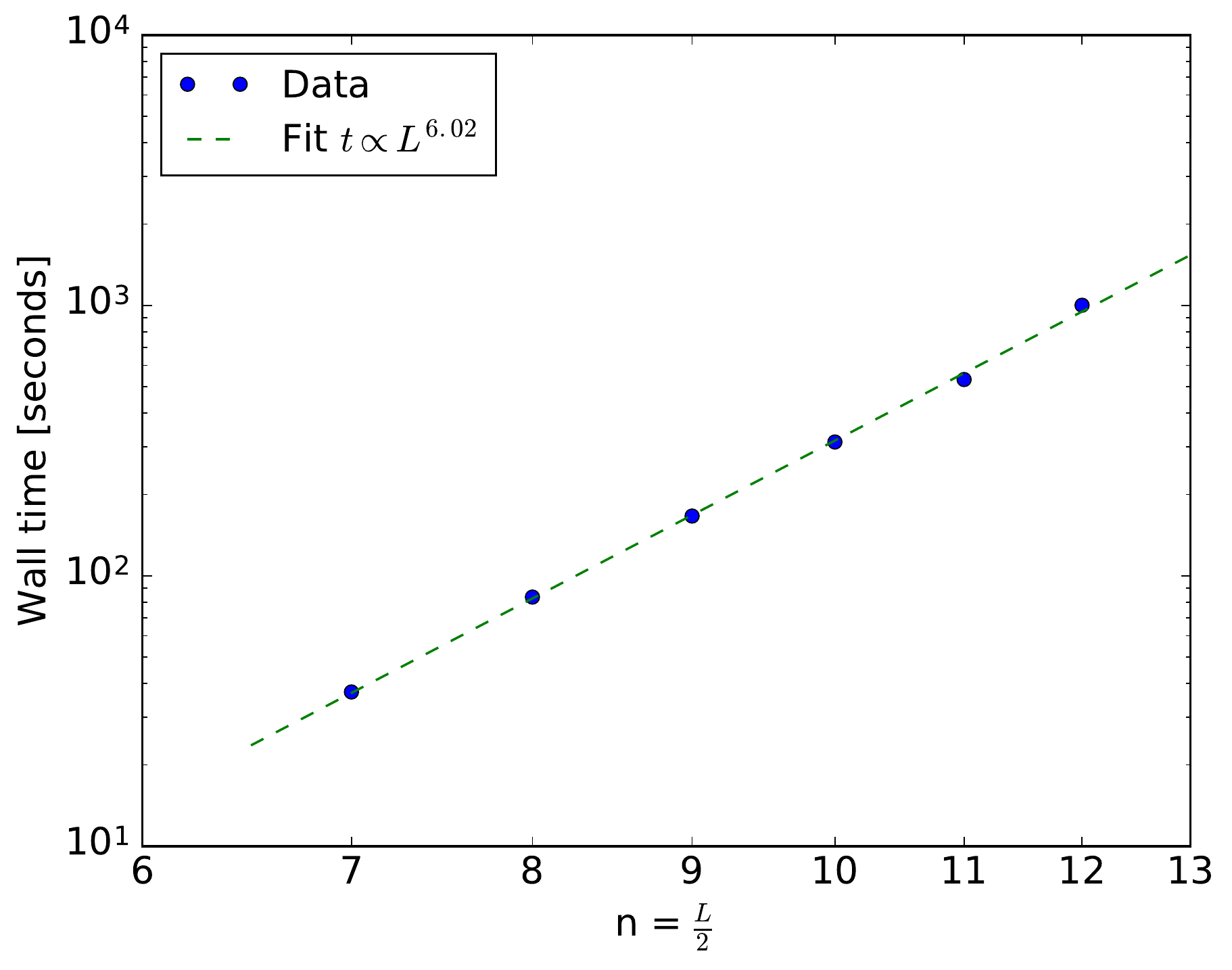}
 \caption{\label{timings_figure} Scaling of the wall time to calculate the 3-RDM for the ($2n$, $2n$) active spaces of all-trans polyenes C$_{2n}$H$_{2n+2}$ for a reduced virtual dimension $D=1000$. All computations were performed with \textsc{chemps2} on a single node with a dual Intel Xeon Sandy Bridge E5-2670 (total of 16 cores at 2.6 GHz) and 64 GB of memory. The geometry of the polyenes is optimized with B3LYP/cc-pVDZ. The active spaces are fully converged with DMRG-SCF($2n$, $2n$)/cc-pVDZ. Localized orbitals are constructed with the Edmiston-Ruedenberg algorithm, and ordered according to the topology of the molecule with the Fiedler vector of the exchange matrix. The scaling of the wall time has to be compared with the theoretical computational scaling $\mathcal{O}(L^{4} D^3 + L^{6} D^2)$.}
\end{figure}

The 2-, 3-, and 4-particle reduced density matrices (2-, 3-, and 4-RDM) are defined as
\begin{eqnarray}
 \Gamma^2_{pq;wx} & = & \sum\limits_{\sigma \tau} \braket{ \hat{a}_{p\sigma}^{\dagger} \hat{a}_{q\tau}^{\dagger} \hat{a}_{x\tau} \hat{a}_{w\sigma} },\\
 \Gamma^3_{pqr; wxy} & = & \sum\limits_{ \sigma \tau \upsilon } \braket{ \hat{a}_{p\sigma}^{\dagger} \hat{a}_{q\tau}^{\dagger} \hat{a}_{r\upsilon}^{\dagger} \hat{a}_{y\upsilon} \hat{a}_{x\tau} \hat{a}_{w\sigma} },\\
 \Gamma^4_{pqrs; wxyz} & = & \sum\limits_{ \sigma \tau \upsilon \phi } \braket{ \hat{a}_{p\sigma}^{\dagger} \hat{a}_{q\tau}^{\dagger} \hat{a}_{r\upsilon}^{\dagger} \hat{a}_{s\phi}^{\dagger} \hat{a}_{z\phi} \hat{a}_{y\upsilon} \hat{a}_{x\tau} \hat{a}_{w\sigma} },
\end{eqnarray}
where the Greek letters denote spin-projections and the Latin letters spatial orbitals. The computational cost to obtain the N-RDM as the expectation value of an MPS is $\mathcal{O}(L^{N+1} D^3 + L^{2N} D^2)$.\cite{JCP.128.144115, JCP.128.144117, JCP.135.094104, JCP.141.174111, JCTC.12.1583} In \textsc{chemps2}, the 2- and 3-RDM are implemented as described in these works. Renormalized operators of three spin-$\frac{1}{2}$ second quantized operators \eqref{operator_dagger}-\eqref{operator_nodagger} are needed for the 3-RDM. They couple to two spin-$\frac{1}{2}$ and one spin-$\frac{3}{2}$ renormalized operators:
\begin{equation}
\frac{1}{2} \otimes \frac{1}{2} \otimes \frac{1}{2} = \frac{1}{2} \oplus \frac{1}{2} \oplus \frac{3}{2}.
\end{equation}
The full twelvefold permutation symmetry of the 3-RDM is taken into account, as well as the Hermitian conjugation equivalence of renormalized operators\cite{CPC.185.1501} and the fermionic anti-commutation relations. The scaling of the computational cost of the 3-RDM with system size is illustrated in Fig.~\ref{timings_figure}. The caption of Fig.~\ref{timings_figure} contains all details of the calculation. As can be observed, $\mathcal{O}(L^6 D^2)$ is the dominant contribution in practice.

The generalized Fock matrix will be introduced in Sect.~\ref{section_dmrg_active_space_solver}. This symmetric matrix has two spatial orbital indices and is diagonal in the irreducible representations:
\begin{equation}
   F_{pq} = F_{qp} = F_{pq} \delta_{I_p,I_q}.
\end{equation}
The contraction of the generalized Fock matrix with the 4-RDM of the active space is required for CASPT2:
\begin{equation}
   \left( F.\Gamma^4 \right)_{pqr; wxy} = \sum\limits_{sz} F_{sz} \Gamma^4_{pqrs; wxyz}. \label{f_4rdm_contraction}
\end{equation}
One way to avoid the implementation and computation of the full 4-RDM is to work in the pseudocanonical orbital basis which diagonalizes the generalized Fock matrix:
\begin{equation}
   F_{pq} =  F_{pp} \delta_{p,q}.
\end{equation}
Kurashige and Yanai describe an efficient contraction of the 4-RDM with the \textit{pseudocanonical} Fock matrix, in which two of the eight 4-RDM indices are always identical.\cite{JCP.135.094104}
This is in general a good strategy for smaller active spaces, or molecules with a large point group. For all-trans polyenes, however, it is better to use a localized and ordered orbital basis because the reduced virtual dimension $D$ can then be orders of magnitude smaller.

Another option to avoid the implementation and computation of the full 4-RDM is to use a cumulant approximation.\cite{JCP.141.174111}
With this approximation, imaginary level shifts become necessary, and even then potential energy surfaces can become quite rugged.\cite{JCP.141.174111}
In order to compute the longitudinal static second hyperpolarizability, accurate finite differences need to be calculated, and we have observed that the cumulant approximation of the 4-RDM yields meaningless results for this purpose.

We adopt another strategy to avoid the implementation of the full 4-RDM. Because the generalized Fock matrix is symmetric, the following sum of 4-RDM elements can be used as well:
\begin{equation}
\Gamma^4_{pqrs; wxyz} + \Gamma^4_{pqrz; wxys},
\end{equation}
where the spatial orbitals $s$ and $z$ belong to the same irreducible representation ($I_s = I_z$).
With $\ket{\Psi_0}$ the reference wavefunction and $\hat{E}_{pq} = \sum_{\sigma} \hat{a}^{\dagger}_{p \sigma} \hat{a}_{q\sigma}$, we define `excited' wavefunctions as:
\begin{equation}
\ket{ sz, \alpha, \beta } = \left[ \alpha \left( \hat{E}_{sz} + {E}_{zs} \right) + \beta \right] \ket{\Psi_0},
\end{equation}
which have the same symmetry as the reference wavefunction if $I_s = I_z$.
We denote the 3-RDM of the (unnormalized) excited wavefunctions as
\begin{multline}
  \Gamma(sz,\alpha,\beta)^3_{pqr; wxy} = \\ \sum\limits_{ \sigma \tau \upsilon } \braket{sz, \alpha, \beta \mid \hat{a}_{p\sigma}^{\dagger} \hat{a}_{q\tau}^{\dagger} \hat{a}_{r\upsilon}^{\dagger} \hat{a}_{y\upsilon} \hat{a}_{x\tau} \hat{a}_{w\sigma} \mid sz, \alpha, \beta }.
\end{multline}
In this notation $\Gamma( sz, 0, 1 )^3_{pqr; wxy} = \Gamma^3_{pqr; wxy}$, the 3-RDM of the reference wavefunction.
The following identity holds:
\begin{eqnarray}
 & & 2 \left[ \Gamma^4_{pqrs; wxyz} + \Gamma^4_{pqrz; wxys} \right] + \Gamma^3_{pqr; wxy} \nonumber \\
 & = & \Gamma( sz, 1, 1 )^3_{pqr; wxy} - \Gamma( sz, 1, 0 )^3_{pqr; wxy} \nonumber \\
 & - & \delta_{s,p} \Gamma^3_{zqr; wxy} - \delta_{s,q} \Gamma^3_{pzr; wxy} - \delta_{s,r} \Gamma^3_{pqz; wxy} \nonumber \\
 & - & \delta_{s,w} \Gamma^3_{pqr; zxy} - \delta_{s,x} \Gamma^3_{pqr; wzy} - \delta_{s,y} \Gamma^3_{pqr; wxz} \nonumber \\
 & - & \delta_{z,p} \Gamma^3_{sqr; wxy} - \delta_{z,q} \Gamma^3_{psr; wxy} - \delta_{z,r} \Gamma^3_{pqs; wxy} \nonumber \\
 & - & \delta_{z,w} \Gamma^3_{pqr; sxy} - \delta_{z,x} \Gamma^3_{pqr; wsy} - \delta_{z,y} \Gamma^3_{pqr; wxs}. \label{big_4rdm_formula}
\end{eqnarray}
Eq.~\eqref{big_4rdm_formula} shows that the required 4-RDM contributions can be obtained with $\mathcal{O}(L^2)$ 3-RDM calculations, with total computational cost $\mathcal{O}(L^{6} D^3 + L^{8} D^2)$. This is higher than the theoretical optimum $\mathcal{O}(L^{5} D^3 + L^{8} D^2)$ for the 4-RDM, but as illustrated in Fig.~\ref{timings_figure} for the 3-RDM, the dominant contribution for our 4-RDM calculation will be $\mathcal{O}(L^2 \times L^6 D^2)$.

The excited wavefunctions $\ket{\Psi_1} = \ket{ sz, \alpha, \beta }$ are decomposed into spin-adapted MPS in \textsc{chemps2} by minimization of the Hylleraas functional
\begin{equation}
\mathcal{L} = \braket{ \Psi_1 \mid \Psi_1 } - 2 \braket{ \Psi_1 \mid \alpha \left( \hat{E}_{sz} + {E}_{zs} \right) + \beta \mid \Psi_0 },
\end{equation}
in a manner entirely similar to Ref.~\onlinecite{JCP.141.111101}. The sweep algorithm is only performed between orbitals $\min(s,z)$ and $\max(s,z)$ as the MPS tensors outside of this range do not change. The cost of one full optimization is $\mathcal{O}( L D^2 + L D^3 ) = \mathcal{O}( L D^3 )$, i.e. entirely negligible compared to the subsequent 3-RDM calculation.
After submission of our manuscript, a similar strategy in the \textsc{block} code to compress the perturber wavefunctions for DMRG-NEVPT2 has been brought to our attention.\cite{JCTC.12.1583, perturber-page}

The proposed strategy resembles the strategy in a FCI code to calculate the N-RDM, which is driven by a routine to compute $\ket{\Psi_{pw; \alpha}} = \hat{E}_{pw} \ket{ \Psi_{\alpha} }$. The computation of the 2-RDM, for example, is realized by a nested for-loop over single-particle excitations which generates
\begin{equation}
\ket{\Psi_{pw; qx}} = \hat{E}_{pw} \ket{ \Psi_{qx} } = \hat{E}_{pw} \hat{E}_{qx} \ket{ \Psi_0 }.
\end{equation}
The 2-RDM elements are obtained by taking overlaps with the reference wavefunction:
\begin{equation}
\Gamma^2_{pq, wx} = \braket{ \Psi_0 \mid \Psi_{pw; qx} } - \delta_{q,w} \braket{ \Psi_0 \mid \Psi_{px}  }.
\end{equation}

\section{DMRG-CASPT2} \label{section_dmrg_active_space_solver}
In the complete active space self consistent field method (CASSCF), the spatial orbitals are divided into core, active, and virtual orbitals. The core orbitals are doubly occupied, while the virtual orbitals remain empty. By taking the Coulomb and exchange interactions with the electrons in the core orbitals into account, an effective active space Hamiltonian can be constructed, and its desired eigenstate can be computed with FCI. The gradient and Hessian of the energy with respect to rotations between the three orbital spaces can be computed based on the 2-RDM of the active space solution.\cite{JCP.74.2384} The three orbital spaces are then optimized with the Newton-Raphson algorithm, or its augmented Hessian variant.\cite{JPC.89.52} An important question is the selection of the active orbital space. We want to mention Refs.~\onlinecite{JCP.140.241103}, \onlinecite{JCP.142.244104}, and \onlinecite{JCTC.12.1760}, which shed new light on this subject.

The equations in Ref.~\onlinecite{JCP.74.2384} depend solely on the active space 2-RDM, and any method which can compute this quantity to high accuracy can be used as active space solver. Recently, unbiased RDMs have been obtained with FCI quantum Monte Carlo (FCIQMC),\cite{JCP.141.244117} and a corresponding FCIQMC-CASSCF algorithm was developed.\cite{JCTC.12.1245} We also want to mention a CASSCF variant without underlying wavefunction ansatz, based on the variational optimization of the 2-RDM.\cite{JCTC.v2dm} With DMRG as active space solver, the method is called DMRG-CASSCF or DMRG-SCF.\cite{JCP.128.144116, JCP.128.144117, IJQC.109.2178, JCP.140.241103}

Systems exhibit static correlation when multiple Slater determinants are required for a qualitatively accurate description. In quantum chemistry, the set of important orbitals, in which the occupation changes over the dominant Slater determinants, is typically small. These orbitals form the actice space, and the static correlation can be resolved with CASSCF, FCIQMC-CASSCF, or DMRG-SCF. Due to the Coulomb repulsion between the electrons, the core and virtual orbitals show small deviations from doubly occupied and empty, respectively. The associated energy contribution is called the dynamic correlation. With DMRG as active space solver, it can be resolved with canonical transformation theory,\cite{JCP.132.024105, PCCP.14.7809} internally contracted perturbation theory,\cite{JCP.135.094104, JCP.141.174111, JCTC.12.1583} or a configuration interaction expansion.\cite{JCP.139.044118, JCTC.11.5120} Alternatively, the perturbation wavefunctions can also be solved within the DMRG framework.\cite{JCP.141.111101, JCP.143.102815, JCP.144.034103}

We have implemented internally contracted complete active space second order perturbation theory (CASPT2)\cite{JPC.94.5483, JCP.96.1218} with DMRG as active space solver.\cite{JCP.135.094104} It is based on the generalized Fock operator:
\begin{equation}
 \hat{F} = \sum\limits_{pq} F_{pq} \hat{E}_{pq},
\end{equation}
with matrix elements
\begin{eqnarray}
F_{pq} & = & \frac{1}{2} \sum\limits_{\sigma} \braket{ \hat{a}_{p\sigma} \left[ \hat{H}, \hat{a}_{q \sigma}^{\dagger} \right] - \hat{a}_{p\sigma}^{\dagger} \left[ \hat{H}, \hat{a}_{q \sigma} \right]  } \\
       & = & t_{pq} + \sum\limits_{rs} \braket{ \hat{E}_{rs} } \left( \left( pq | rs \right) - \frac{1}{2} \left( pr | qs \right) \right),
\end{eqnarray}
where $t_{pq}$ and $(pq|rs)$ are the usual one- and two-electron integrals.

The full Hilbert space $\mathcal{H}$ is split up into four parts:
\begin{equation}
 \mathcal{H} = \mathcal{V}_0 \oplus \mathcal{V}_{\text{K}} \oplus \mathcal{V}_{\text{SD}} \oplus \mathcal{V}_{\text{TQ..}}.
\end{equation}
$\mathcal{V}_0$ contains only the CASSCF solution $\ket{\Psi_0}$. $\mathcal{V}_{\text{K}}$ is the space spanned by all possible active space excitations on top of $\ket{\Psi_0}$ which are orthogonal to $\mathcal{V}_0$. Wavefunctions in $\mathcal{V}_{\text{K}}$ have the same core and virtual orbitals as $\ket{\Psi_0}$, with the same occupation. $\mathcal{V}_{{\text{SD}}}$ contains all single and double particle excitations on top of $\ket{\Psi_0}$ which are orthogonal to $\mathcal{V}_0 \oplus \mathcal{V}_{\text{K}}$. With the indices $ij$ for core orbitals, $tuv$ for active orbitals, and $ab$ for virtual orbitals, $\mathcal{V}_{{\text{SD}}}$ is spanned by the following excitation types:
\begin{eqnarray}
\text{A} & : & \quad \hat{E}_{ti} \hat{E}_{uv} \ket{\Psi_0}, \label{excitation_A}\\
\text{B} & : & \quad \hat{E}_{ti} \hat{E}_{uj} \ket{\Psi_0}, \\
\text{C} & : & \quad \hat{E}_{at} \hat{E}_{uv} \ket{\Psi_0}, \\
\text{D} & : & \quad \hat{E}_{ai} \hat{E}_{tu} \ket{\Psi_0},~\hat{E}_{ti}\hat{E}_{au}\ket{\Psi_0}, \\
\text{E} & : & \quad \hat{E}_{ti} \hat{E}_{aj} \ket{\Psi_0}, \\
\text{F} & : & \quad \hat{E}_{at} \hat{E}_{bu} \ket{\Psi_0}, \\
\text{G} & : & \quad \hat{E}_{ai} \hat{E}_{bt} \ket{\Psi_0}, \\
\text{H} & : & \quad \hat{E}_{ai} \hat{E}_{bj} \ket{\Psi_0}.\label{excitation_H}
\end{eqnarray}
And $\mathcal{V}_{{\text{TQ..}}}$ is the remainder of $\mathcal{H}$.
The zeroth order Hamiltonian for internally contracted CASPT2 is
\begin{equation}
 \hat{H}_0 = \hat{P}_0 \hat{F} \hat{P}_0 + \hat{P}_{\text{K}} \hat{F} \hat{P}_{\text{K}} + \hat{P}_{{\text{SD}}} \hat{F} \hat{P}_{{\text{SD}}} + \hat{P}_{{\text{TQ..}}} \hat{F} \hat{P}_{{\text{TQ..}}},
\end{equation}
where $\hat{P}_{\text{X}}$ is the projector onto $\mathcal{V}_{\text{X}}$.
The first order wavefunction $\ket{\Psi_1}$ for internally contracted CASPT2 is spanned by a linear combination over $\mathcal{V}_{\text{SD}}$:
\begin{eqnarray}
 \ket{\Psi_1} & = & \sum_{pq;rs \in \mathcal{V}_{\text{SD}}} C_{pq;rs} \hat{E}_{pq} \hat{E}_{rs} \ket{\Psi_0} \nonumber \\
              & = & \sum_{pq;rs \in \mathcal{V}_{\text{SD}}} C_{pq;rs} \ket{ \Psi_{pq;rs} }.
\end{eqnarray}
The coefficients can be found by solving
\begin{multline}
 \sum_{pq;rs \in \mathcal{V}_{\text{SD}}} \braket{ \Psi_{wx;yz} \mid \hat{H}_0 - E_0 \mid \Psi_{pq;rs} } C_{pq;rs} \\ = - \braket{ \Psi_{wx;yz} \mid \hat{H} \mid \Psi_0}. \label{caspt2_equation}
\end{multline}
The overlap matrix $\braket{ \Psi_{wx;yz} \mid \Psi_{pq;rs} } $ is block-diagonal in the different excitation types (A to H). It is diagonalized, small eigenvalues are discarded, and Eq.~\eqref{caspt2_equation} is transformed to
\begin{equation}
\sum\limits_{ \beta } \left( \mathcal{F}_{\alpha\beta} - E_0 \delta_{\alpha,\beta} \right) \mathcal{C}_{\beta} = - \mathcal{V}_{\alpha}, \label{equation_in_practice_caspt2}
\end{equation}
with $\mathcal{F}$ diagonal for two excitations $\ket{\alpha}$ and $\ket{\beta}$ of the same type (A to H). In \textsc{chemps2}, the coefficients $\mathcal{C}_{\alpha}$ are solved with the conjugate gradient algorithm. We use the initial guess
\begin{equation}
\mathcal{C}_{\alpha}^{\text{ini}} = - \frac{ \mathcal{V}_{\alpha} }{ \mathcal{F}_{\alpha\alpha} - E_0 }.
\end{equation}

For the excitation types A and C, the contraction of the generalized Fock matrix with 4-RDM of the active space of the CASSCF solution is needed. If the active space orbitals in the DMRG algorithm are not pseudocanonical, we rotate $\Gamma^1$, $\Gamma^2$, $\Gamma^3$, and $\left(F.\Gamma^4\right)$ to the pseudocanonical orbital basis before building the required intermediates to solve Eq.~\eqref{equation_in_practice_caspt2}. We also couple the excitation types B, E, F, G, and H to singlet and triplet excitations:
\begin{eqnarray}
\text{B singlet} & : & \quad \left( \hat{E}_{ti} \hat{E}_{uj} + \hat{E}_{tj} \hat{E}_{ui} \right) \ket{\Psi_0}, \\
\text{B triplet} & : & \quad \left( \hat{E}_{ti} \hat{E}_{uj} - \hat{E}_{tj} \hat{E}_{ui} \right) \ket{\Psi_0},
\end{eqnarray}
to make $\mathcal{F}$ more sparse.\cite{JPC.94.5483, JCP.96.1218}

In order to mitigate intruder state problems, we have implemented the imaginary level shift\cite{CPL.274.196} and the ionization potential - electron affinity (IPEA) shift.\cite{CPL.396.142} For the latter, the matrix on the left-hand side in Eq.~\eqref{caspt2_equation} is shifted with
\begin{multline}
\braket{\Psi_{ wx;yz } \mid \hat{F} \mid \Psi_{pq;rs }} \mathrel{+}= \\
\delta_{p,w} \delta_{q,x} \delta_{r,y} \delta_{s,z} \frac{ \epsilon^{\text{IPEA}}}{2} \braket{ \Psi_{wx;yz} \mid \Psi_{pq;rs} } \\ \times \left( 4 + \braket{ \hat{E}_{pp} } - \braket{ \hat{E}_{qq} } + \braket{ \hat{E}_{rr} } - \braket{ \hat{E}_{ss} } \right).
\end{multline}

\section{Longitudinal static second hyperpolarizability of polyenes} \label{gamma_for_polyenes}
Static second hyperpolarizabilities can be obtained with the finite field method.\cite{JCC.11.82} When an electric field $\mathcal{E}$ is applied in the $z$-direction, the Hamiltonian becomes
\begin{equation}
\hat{H} = \hat{H}_0 + \mathcal{E}z.
\end{equation}
The static second hyperpolarizability in the $z$-direction can be calculated as the fourth order derivative of the eigenvalue $E$ of $\hat{H}$ with respect to the electric field:
\begin{equation}
 \gamma_{zzzz} = - \left( \frac{ \partial^4 E }{ \partial \mathcal{E}^4 } \right)_{\mathcal{E} \rightarrow 0}.
\end{equation}
For centrosymmetric all-trans polyenes, the fourth order derivative can be approximated with the finite difference
\begin{equation}
 \gamma_{zzzz}(\mathcal{E}) = \frac{ -6E(0) + 8E(\mathcal{E} ) - 2E(2\mathcal{E} ) }{ \mathcal{E}^4 },
\end{equation}
when the origin is chosen in the center of the polyene.
We calculate $\gamma_{zzzz}(\mathcal{E})$ for $\mathcal{E} \in \{ 0.001, 0.002, 0.004 \}$ a.u. and extrapolate the finite differences to zero field with a least-squares fit to
\begin{equation}
 \gamma_{zzzz}(\mathcal{E}) = \gamma_{zzzz}( 0 ) + c_1 \mathcal{E}^2 + c_2 \mathcal{E}^4, \label{extrapol_to_zero_field}
\end{equation}
with $c_2 \ll c_1$. The odd powers in the electric field vanish in Eq.~\eqref{extrapol_to_zero_field} because of the centrosymmetry.

The values of $\mathcal{E}$ have to be chosen with care.\cite{JCP.136.134110} For too large fields, higher order effects come into play in Eq.~\eqref{extrapol_to_zero_field}. The \textit{ab initio} calculations yield energies with a certain precision, due to either the convergence threshold or the finite precision arithmetic. For too small fields, the relative error on $\gamma_{zzzz}(\mathcal{E})$ will therefore be too large.

All calculations were performed in the cc-pVDZ basis. The geometries of all-trans polyenes C$_{2n}$H$_{2n+2}$ [$n=4-12$] with $\mathsf{C_{2h}}$ symmetry were optimized with the B3LYP functional in \textsc{psi4}.\cite{WIRCMS.2.556} The carbon atoms of the polyenes form two parallel rows, and the electric fields have the same direction. The electric fields break the $\mathsf{C_{2h}}$ symmetry to $\mathsf{C_s}$ symmetry, and the latter was used in the \textit{ab initio} calculations. The RHF, MP2, and CCSD calculations were performed with \textsc{pyscf},\cite{pyscf_github} as well as the calculation of the electron integrals for \textsc{chemps2}. For the DMRG-SCF and DMRG-CASPT2 calculations, a $(2n,2n)$ $\pi$-electron active space was used. The active space orbitals were localized with an augmented Hessian Newton-Raphson implementation of the Edmiston-Ruedenberg algorithm.\cite{RevModPhys.35.457, CPC.191.235} They were ordered according to the one-dimensional topology of the polyene, by means of the Fiedler vector of the exchange matrix.\cite{PhysRevA.83.012508, CPC.191.235} The DMRG calculations were performed with a reduced virtual dimension $D=1000$, and a residual norm threshold $10^{-10}$ for the Davidson algorithm. With these parameters, both the ground state $\ket{\Psi_0}$ of the active space Hamiltonian and the corresponding excited wavefunctions $\ket{sz,\alpha,\beta}$ are indistinguishable from FCI.

In order to obtain accurate finite differences $\gamma_{zzzz}(\mathcal{E})$ and corresponding extrapolations \eqref{extrapol_to_zero_field} to zero field, we have observed that the cumulant approximation, the imaginary level shift, and the IPEA shift cannot be used. The IPEA shift even yields negative $\gamma_{zzzz}$.

\begin{figure*}[t]
 \includegraphics[width=0.93\textwidth]{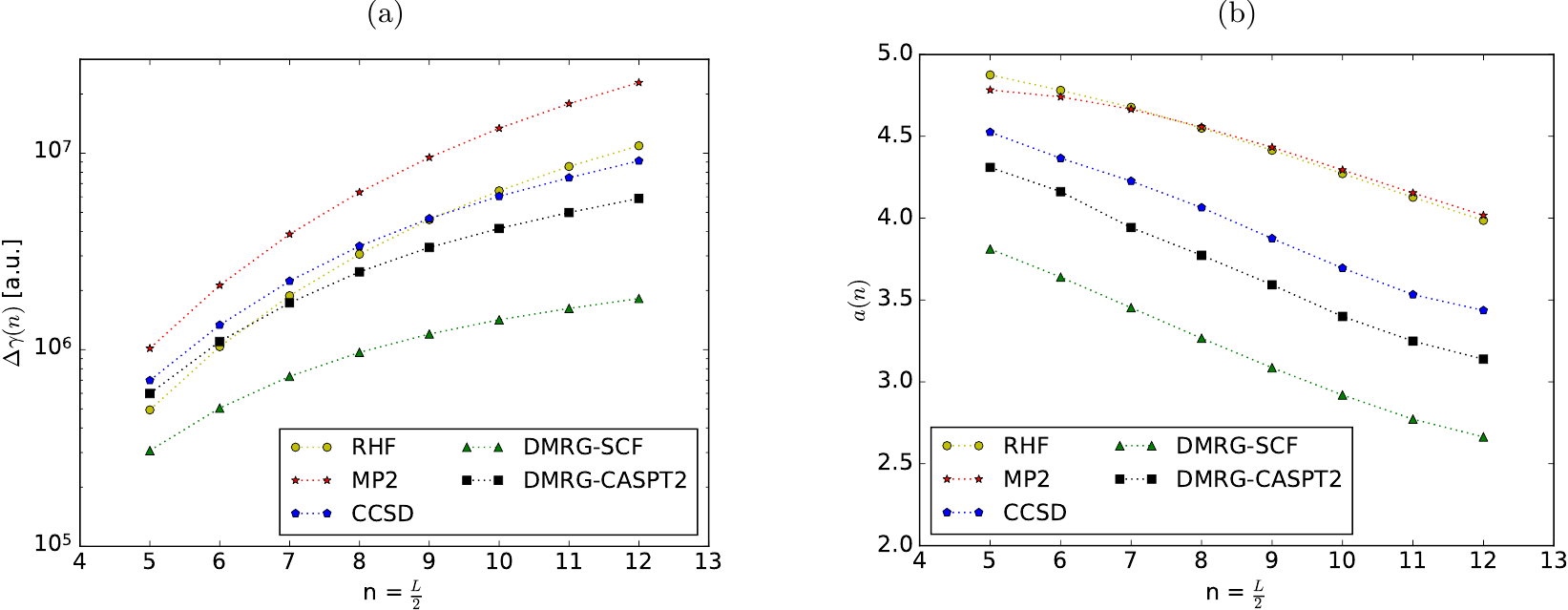}
 \caption{\label{hyperpol_figure} The longitudinal static second hyperpolarizability of all-trans polyenes C$_{2n}$H$_{2n+2}$. The geometries are optimized with B3LYP/cc-pVDZ and all calculations are performed with the cc-pVDZ basis. (a) The incremental longitudinal static second hyperpolarizability $\Delta \gamma_{zzzz}(n) = \gamma_{zzzz}(n) - \gamma_{zzzz}(n-1)$. (b) The estimatation in Eq.~\eqref{power_hyperpol} for the power law exponent. }
\end{figure*}

The power law behaviour $\gamma_{zzzz}(n) \propto n^{a(n)}$ is often assumed.\cite{CR.94.243} Consider a small local electric field which causes a response of a certain length scale. For polyenes smaller than this length scale, the possibility for response opens up with polyene length $n$, and a rapid increase of $\gamma_{zzzz}$ with $n$ is observed ($a(n) \gg 1$). For polyenes much larger than this length scale, $\gamma_{zzzz}$ scales linearly with $n$ ($a(n) \rightarrow 1$). The incremental longitudinal static second hyperpolarizability
\begin{equation}
\Delta \gamma_{zzzz}(n) = \gamma_{zzzz}(n) - \gamma_{zzzz}(n-1), \label{incremental_hyperpol}
\end{equation}
and the estimated exponent for power law behaviour
\begin{equation}
a(n) \approx \frac{ \log( \gamma_{zzzz}(n) ) - \log( \gamma_{zzzz}(n - 1) ) }{ \log( n ) - \log( n - 1 ) }, \label{power_hyperpol}
\end{equation}
are shown in Figs.~\ref{hyperpol_figure}(a) and \ref{hyperpol_figure}(b), respectively.

\begin{figure}[t]
 \includegraphics[width=0.465\textwidth]{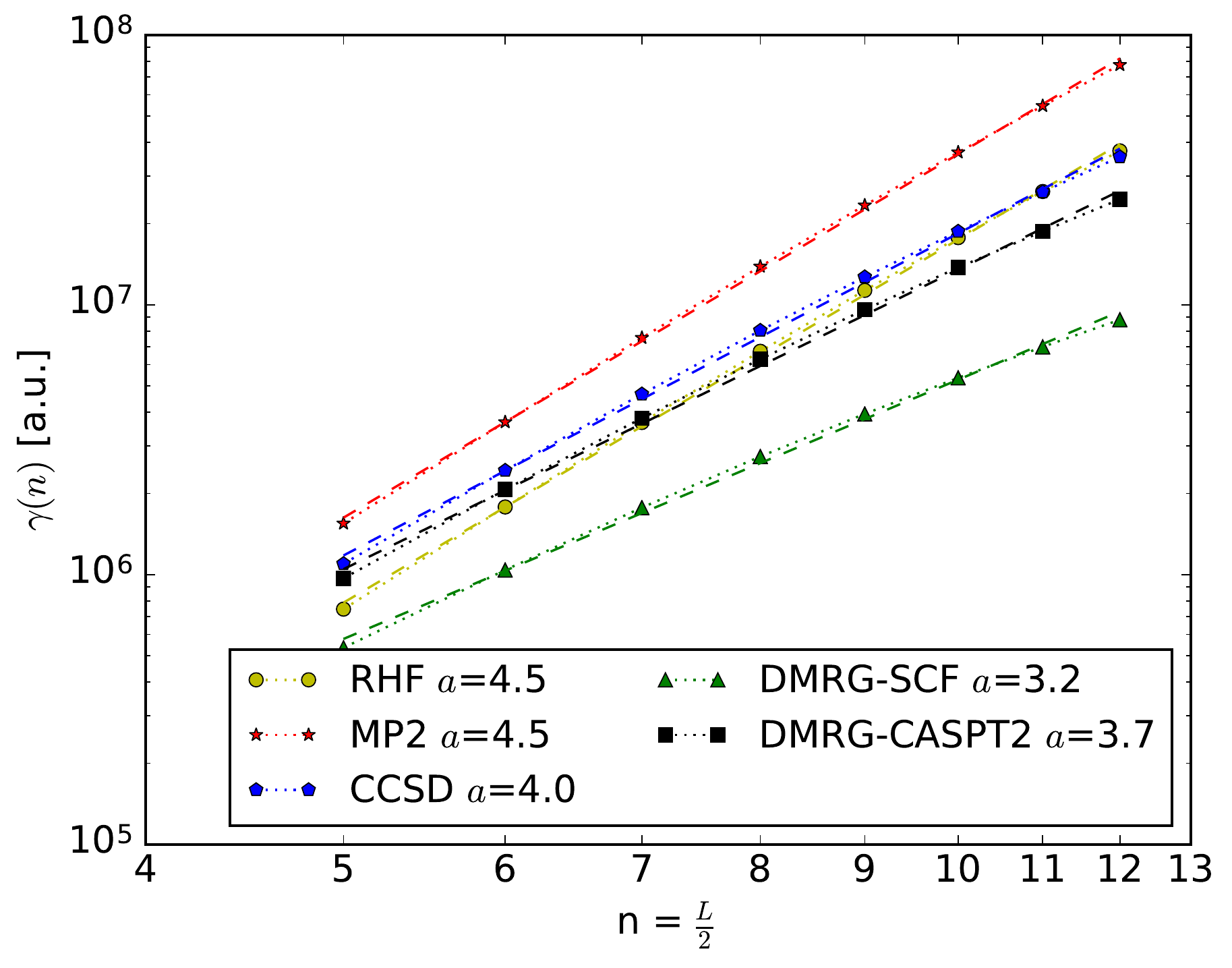}
 \caption{\label{power-law-fit} The power law scaling of the longitudinal static second hyperpolarizability of all-trans polyenes C$_{2n}$H$_{2n+2}$ is determined with a linear fit on a double logarithmic scale. }
\end{figure}

Of all methods considered, DMRG-SCF and DMRG-CASPT2 yield the lowest incremental longitudinal static second hyperpolarizabilities $\Delta \gamma_{zzzz}(n)$ and exponents $a(n)$. An experimental value of $a = 3.2$ was determined for polyenes with length $n=4-16$, with a linear fit on a double logarithmic scale.\cite{JACS.115.860} We have performed a similar analysis for our computational data in Fig.~\ref{power-law-fit}. RHF, MP2, CCSD, DMRG-SCF, and DMRG-CASPT2 yield the exponents $a = 4.5$, $a=4.5$, $a=4.0$, $a = 3.2$, and $a=3.7$, respectively. While the DMRG-SCF exponent corresponds best to the experimental result, the calculations have been performed in vacuum and with a modest basis set, and should therefore be treated with care.

The \textit{ab initio} calculations mainly allow to compare different levels of theory. As noted in the introduction, CCSD yields larger (smaller) longitudinal static second hyperpolarizabilities than RHF for short (long) polyenes. The RHF and MP2 values and power law exponents are substantially reduced with DMRG-SCF and DMRG-CASPT2, respectively. Our calculations hence point out the importance of static correlation for the non-linear optical properties of conjugated molecules.

The CCSD and DMRG-CASPT2 (incremental) $\gamma_{zzzz}$ differ by less than a factor 2. It remains an open question to which extent CCSD covers the static correlation incorporated in DMRG-CASPT2. If CCSD only captures a fraction of the static correlation, multi-reference coupled cluster theory (MRCC) can significantly alter the (incremental) $\gamma_{zzzz}$ compared to DMRG-CASPT2, in analogy to the single-reference calculations.


\section{Summary} \label{section_conclusions}
In Sect.~\ref{section_mps_ansatz} and \ref{section_dmrg_method} the matrix product state (MPS) ansatz and the density matrix renormalization group (DMRG) algorithm were reviewed. A hybrid parallelization of DMRG for mixed distributed and shared memory architectures was also described. Processes become responsible for certain operator and complementary operator pairs, while the contractions over separate reduced symmetry sectors are parallelized by threads. Because the two parallelization strategies are independent, they show independent (multiplicative) speedups.

In Sect.~\ref{our_implementation} our strategy to contract the generalized Fock matrix with the 4-particle reduced density matrix (4-RDM) of the reference wavefunction was explained. The required 4-RDM elements can be obtained from the 3-RDMs of `excited' wavefunctions, formed by symmetry-conserving single-particle excitations on top of the reference wavefunction. These excited wavefunctions are decomposed into spin-adapted MPSs at negligible cost. The total computational cost of our strategy scales as $\mathcal{O}(L^6D^3 + L^8D^2)$. In practice, the dominant term is $\mathcal{O}(L^8D^2)$, which is the same as for the theoretical optimum $\mathcal{O}(L^5D^3 + L^8D^2)$.

Our implementation of DMRG-CASPT2 was outlined in Sect.~\ref{section_dmrg_active_space_solver}. We have studied the longitudinal static second hyperpolarizability of all-trans polyenes C$_{2n}$H$_{2n+2}$ [$n=4-12$], obtained in the cc-pVDZ basis with the finite field method, in Sect.~\ref{gamma_for_polyenes}. The results of three single-reference methods (RHF, MP2, and CCSD) were compared with the results of DMRG-SCF and DMRG-CASPT2, using a $(2n,2n)$ $\pi$-electron active space. The multi-reference methods yield substantially lower values and exponents for the longitudinal static second hyperpolarizability than their single-reference counterparts.
Our calculations hence point out the importance of static correlation for the non-linear optical properties of conjugated molecules.

\section*{Acknowledgements}
S.W. would like to thank Ward Poelmans and Ewan Higgs for their help with improving the disk bandwidths in \textsc{chemps2}; Jun Yang and Sandeep Sharma for their help with the installation and execution of \textsc{block}; Toru Shiozaki and Takeshi Yanai for insightful conversations on CASPT2; and Peter Limacher for stimulating discussions on the second hyperpolarizability. S.W. also gratefully acknowledges a postdoctoral fellowship from the Research Foundation Flanders (Fonds Wetenschappelijk Onderzoek Vlaanderen). The computational resources (Stevin Supercomputer Infrastructure) and services used in this work were provided by the VSC (Flemish Supercomputer Center), funded by Ghent University, the Hercules Foundation and the Flemish Government - department EWI.

\bibliographystyle{aipnum4-1}
\bibliography{biblio}

\begin{thebibliography}{80}%
\makeatletter
\providecommand \@ifxundefined [1]{%
 \@ifx{#1\undefined}
}%
\providecommand \@ifnum [1]{%
 \ifnum #1\expandafter \@firstoftwo
 \else \expandafter \@secondoftwo
 \fi
}%
\providecommand \@ifx [1]{%
 \ifx #1\expandafter \@firstoftwo
 \else \expandafter \@secondoftwo
 \fi
}%
\providecommand \natexlab [1]{#1}%
\providecommand \enquote  [1]{``#1''}%
\providecommand \bibnamefont  [1]{#1}%
\providecommand \bibfnamefont [1]{#1}%
\providecommand \citenamefont [1]{#1}%
\providecommand \href@noop [0]{\@secondoftwo}%
\providecommand \href [0]{\begingroup \@sanitize@url \@href}%
\providecommand \@href[1]{\@@startlink{#1}\@@href}%
\providecommand \@@href[1]{\endgroup#1\@@endlink}%
\providecommand \@sanitize@url [0]{\catcode `\\12\catcode `\$12\catcode
  `\&12\catcode `\#12\catcode `\^12\catcode `\_12\catcode `\%12\relax}%
\providecommand \@@startlink[1]{}%
\providecommand \@@endlink[0]{}%
\providecommand \url  [0]{\begingroup\@sanitize@url \@url }%
\providecommand \@url [1]{\endgroup\@href {#1}{\urlprefix }}%
\providecommand \urlprefix  [0]{URL }%
\providecommand \Eprint [0]{\href }%
\providecommand \doibase [0]{http://dx.doi.org/}%
\providecommand \selectlanguage [0]{\@gobble}%
\providecommand \bibinfo  [0]{\@secondoftwo}%
\providecommand \bibfield  [0]{\@secondoftwo}%
\providecommand \translation [1]{[#1]}%
\providecommand \BibitemOpen [0]{}%
\providecommand \bibitemStop [0]{}%
\providecommand \bibitemNoStop [0]{.\EOS\space}%
\providecommand \EOS [0]{\spacefactor3000\relax}%
\providecommand \BibitemShut  [1]{\csname bibitem#1\endcsname}%
\let\auto@bib@innerbib\@empty
\bibitem [{\citenamefont {Champagne}\ \emph {et~al.}(1998)\citenamefont
  {Champagne}, \citenamefont {Perp\`ete}, \citenamefont {van Gisbergen},
  \citenamefont {Baerends}, \citenamefont {Snijders}, \citenamefont
  {Soubra-Ghaoui}, \citenamefont {Robins},\ and\ \citenamefont
  {Kirtman}}]{JCP.109.10489}%
  \BibitemOpen
  \bibfield  {author} {\bibinfo {author} {\bibfnamefont {B.}~\bibnamefont
  {Champagne}}, \bibinfo {author} {\bibfnamefont {E.~A.}\ \bibnamefont
  {Perp\`ete}}, \bibinfo {author} {\bibfnamefont {S.~J.~A.}\ \bibnamefont {van
  Gisbergen}}, \bibinfo {author} {\bibfnamefont {E.-J.}\ \bibnamefont
  {Baerends}}, \bibinfo {author} {\bibfnamefont {J.~G.}\ \bibnamefont
  {Snijders}}, \bibinfo {author} {\bibfnamefont {C.}~\bibnamefont
  {Soubra-Ghaoui}}, \bibinfo {author} {\bibfnamefont {K.~A.}\ \bibnamefont
  {Robins}}, \ and\ \bibinfo {author} {\bibfnamefont {B.}~\bibnamefont
  {Kirtman}},\ }\href {\doibase 10.1063/1.477731} {\bibfield  {journal}
  {\bibinfo  {journal} {J. Chem. Phys.}\ }\textbf {\bibinfo {volume} {109}},\
  \bibinfo {pages} {10489} (\bibinfo {year} {1998})}\BibitemShut {NoStop}%
\bibitem [{\citenamefont {van Gisbergen}\ \emph {et~al.}(1999)\citenamefont
  {van Gisbergen}, \citenamefont {Schipper}, \citenamefont {Gritsenko},
  \citenamefont {Baerends}, \citenamefont {Snijders}, \citenamefont
  {Champagne},\ and\ \citenamefont {Kirtman}}]{PhysRevLett.83.694}%
  \BibitemOpen
  \bibfield  {author} {\bibinfo {author} {\bibfnamefont {S.~J.~A.}\
  \bibnamefont {van Gisbergen}}, \bibinfo {author} {\bibfnamefont {P.~R.~T.}\
  \bibnamefont {Schipper}}, \bibinfo {author} {\bibfnamefont {O.~V.}\
  \bibnamefont {Gritsenko}}, \bibinfo {author} {\bibfnamefont {E.~J.}\
  \bibnamefont {Baerends}}, \bibinfo {author} {\bibfnamefont {J.~G.}\
  \bibnamefont {Snijders}}, \bibinfo {author} {\bibfnamefont {B.}~\bibnamefont
  {Champagne}}, \ and\ \bibinfo {author} {\bibfnamefont {B.}~\bibnamefont
  {Kirtman}},\ }\href {\doibase 10.1103/PhysRevLett.83.694} {\bibfield
  {journal} {\bibinfo  {journal} {Phys. Rev. Lett.}\ }\textbf {\bibinfo
  {volume} {83}},\ \bibinfo {pages} {694} (\bibinfo {year} {1999})}\BibitemShut
  {NoStop}%
\bibitem [{\citenamefont {Champagne}\ \emph {et~al.}(2000)\citenamefont
  {Champagne}, \citenamefont {Perp\`ete}, \citenamefont {Jacquemin},
  \citenamefont {van Gisbergen}, , \citenamefont {Baerends}, \citenamefont
  {Soubra-Ghaoui}, , \citenamefont {Robins},\ and\ \citenamefont
  {Kirtman}}]{JPCA.104.4755}%
  \BibitemOpen
  \bibfield  {author} {\bibinfo {author} {\bibfnamefont {B.}~\bibnamefont
  {Champagne}}, \bibinfo {author} {\bibfnamefont {E.~A.}\ \bibnamefont
  {Perp\`ete}}, \bibinfo {author} {\bibfnamefont {D.}~\bibnamefont
  {Jacquemin}}, \bibinfo {author} {\bibfnamefont {S.~J.~A.}\ \bibnamefont {van
  Gisbergen}}, , \bibinfo {author} {\bibfnamefont {E.-J.}\ \bibnamefont
  {Baerends}}, \bibinfo {author} {\bibfnamefont {C.}~\bibnamefont
  {Soubra-Ghaoui}}, , \bibinfo {author} {\bibfnamefont {K.~A.}\ \bibnamefont
  {Robins}}, \ and\ \bibinfo {author} {\bibfnamefont {B.}~\bibnamefont
  {Kirtman}},\ }\href {\doibase 10.1021/jp993839d} {\bibfield  {journal}
  {\bibinfo  {journal} {J. Phys. Chem. A}\ }\textbf {\bibinfo {volume} {104}},\
  \bibinfo {pages} {4755} (\bibinfo {year} {2000})}\BibitemShut {NoStop}%
\bibitem [{\citenamefont {Mori-S\'anchez}, \citenamefont {Wu},\ and\
  \citenamefont {Yang}(2003)}]{JCP.119.11001}%
  \BibitemOpen
  \bibfield  {author} {\bibinfo {author} {\bibfnamefont {P.}~\bibnamefont
  {Mori-S\'anchez}}, \bibinfo {author} {\bibfnamefont {Q.}~\bibnamefont {Wu}},
  \ and\ \bibinfo {author} {\bibfnamefont {W.}~\bibnamefont {Yang}},\ }\href
  {\doibase 10.1063/1.1630011} {\bibfield  {journal} {\bibinfo  {journal} {J.
  Chem. Phys.}\ }\textbf {\bibinfo {volume} {119}},\ \bibinfo {pages} {11001}
  (\bibinfo {year} {2003})}\BibitemShut {NoStop}%
\bibitem [{\citenamefont {Kamiya}\ \emph {et~al.}(2005)\citenamefont {Kamiya},
  \citenamefont {Sekino}, \citenamefont {Tsuneda},\ and\ \citenamefont
  {Hirao}}]{JCP.122.234111}%
  \BibitemOpen
  \bibfield  {author} {\bibinfo {author} {\bibfnamefont {M.}~\bibnamefont
  {Kamiya}}, \bibinfo {author} {\bibfnamefont {H.}~\bibnamefont {Sekino}},
  \bibinfo {author} {\bibfnamefont {T.}~\bibnamefont {Tsuneda}}, \ and\
  \bibinfo {author} {\bibfnamefont {K.}~\bibnamefont {Hirao}},\ }\href
  {\doibase 10.1063/1.1935514} {\bibfield  {journal} {\bibinfo  {journal} {J.
  Chem. Phys.}\ }\textbf {\bibinfo {volume} {122}},\ \bibinfo {pages} {234111}
  (\bibinfo {year} {2005})}\BibitemShut {NoStop}%
\bibitem [{\citenamefont {Sekino}\ \emph {et~al.}(2007)\citenamefont {Sekino},
  \citenamefont {Maeda}, \citenamefont {Kamiya},\ and\ \citenamefont
  {Hirao}}]{JCP.126.014107}%
  \BibitemOpen
  \bibfield  {author} {\bibinfo {author} {\bibfnamefont {H.}~\bibnamefont
  {Sekino}}, \bibinfo {author} {\bibfnamefont {Y.}~\bibnamefont {Maeda}},
  \bibinfo {author} {\bibfnamefont {M.}~\bibnamefont {Kamiya}}, \ and\ \bibinfo
  {author} {\bibfnamefont {K.}~\bibnamefont {Hirao}},\ }\href {\doibase
  10.1063/1.2428291} {\bibfield  {journal} {\bibinfo  {journal} {J. Chem.
  Phys.}\ }\textbf {\bibinfo {volume} {126}},\ \bibinfo {pages} {014107}
  (\bibinfo {year} {2007})}\BibitemShut {NoStop}%
\bibitem [{\citenamefont {Song}\ \emph {et~al.}(2008)\citenamefont {Song},
  \citenamefont {Watson}, \citenamefont {Sekino},\ and\ \citenamefont
  {Hirao}}]{JCP.129.024117}%
  \BibitemOpen
  \bibfield  {author} {\bibinfo {author} {\bibfnamefont {J.-W.}\ \bibnamefont
  {Song}}, \bibinfo {author} {\bibfnamefont {M.~A.}\ \bibnamefont {Watson}},
  \bibinfo {author} {\bibfnamefont {H.}~\bibnamefont {Sekino}}, \ and\ \bibinfo
  {author} {\bibfnamefont {K.}~\bibnamefont {Hirao}},\ }\href {\doibase
  10.1063/1.2936830} {\bibfield  {journal} {\bibinfo  {journal} {J. Chem.
  Phys.}\ }\textbf {\bibinfo {volume} {129}},\ \bibinfo {pages} {024117}
  (\bibinfo {year} {2008})}\BibitemShut {NoStop}%
\bibitem [{\citenamefont {Bartlett}\ and\ \citenamefont
  {Purvis}(1979)}]{PhysRevA.20.1313}%
  \BibitemOpen
  \bibfield  {author} {\bibinfo {author} {\bibfnamefont {R.~J.}\ \bibnamefont
  {Bartlett}}\ and\ \bibinfo {author} {\bibfnamefont {G.~D.}\ \bibnamefont
  {Purvis}},\ }\href {\doibase 10.1103/PhysRevA.20.1313} {\bibfield  {journal}
  {\bibinfo  {journal} {Phys. Rev. A}\ }\textbf {\bibinfo {volume} {20}},\
  \bibinfo {pages} {1313} (\bibinfo {year} {1979})}\BibitemShut {NoStop}%
\bibitem [{\citenamefont {Li}\ \emph {et~al.}(2008)\citenamefont {Li},
  \citenamefont {Chen}, \citenamefont {Li},\ and\ \citenamefont
  {Shuai}}]{CPL.457.276}%
  \BibitemOpen
  \bibfield  {author} {\bibinfo {author} {\bibfnamefont {Q.}~\bibnamefont
  {Li}}, \bibinfo {author} {\bibfnamefont {L.}~\bibnamefont {Chen}}, \bibinfo
  {author} {\bibfnamefont {Q.}~\bibnamefont {Li}}, \ and\ \bibinfo {author}
  {\bibfnamefont {Z.}~\bibnamefont {Shuai}},\ }\href {\doibase
  10.1016/j.cplett.2008.04.020} {\bibfield  {journal} {\bibinfo  {journal}
  {Chem. Phys. Lett.}\ }\textbf {\bibinfo {volume} {457}},\ \bibinfo {pages}
  {276} (\bibinfo {year} {2008})}\BibitemShut {NoStop}%
\bibitem [{\citenamefont {Limacher}, \citenamefont {Li},\ and\ \citenamefont
  {L\"uthi}(2011)}]{JCP.135.014111}%
  \BibitemOpen
  \bibfield  {author} {\bibinfo {author} {\bibfnamefont {P.~A.}\ \bibnamefont
  {Limacher}}, \bibinfo {author} {\bibfnamefont {Q.}~\bibnamefont {Li}}, \ and\
  \bibinfo {author} {\bibfnamefont {H.~P.}\ \bibnamefont {L\"uthi}},\ }\href
  {\doibase 10.1063/1.3603967} {\bibfield  {journal} {\bibinfo  {journal} {J.
  Chem. Phys.}\ }\textbf {\bibinfo {volume} {135}},\ \bibinfo {pages} {014111}
  (\bibinfo {year} {2011})}\BibitemShut {NoStop}%
\bibitem [{\citenamefont {Nakano}\ \emph {et~al.}(2012)\citenamefont {Nakano},
  \citenamefont {Minami}, \citenamefont {Fukui}, \citenamefont {Kishi},
  \citenamefont {Shigeta},\ and\ \citenamefont {Champagne}}]{JCP.136.024315}%
  \BibitemOpen
  \bibfield  {author} {\bibinfo {author} {\bibfnamefont {M.}~\bibnamefont
  {Nakano}}, \bibinfo {author} {\bibfnamefont {T.}~\bibnamefont {Minami}},
  \bibinfo {author} {\bibfnamefont {H.}~\bibnamefont {Fukui}}, \bibinfo
  {author} {\bibfnamefont {R.}~\bibnamefont {Kishi}}, \bibinfo {author}
  {\bibfnamefont {Y.}~\bibnamefont {Shigeta}}, \ and\ \bibinfo {author}
  {\bibfnamefont {B.}~\bibnamefont {Champagne}},\ }\href {\doibase
  10.1063/1.3675684} {\bibfield  {journal} {\bibinfo  {journal} {J. Chem.
  Phys.}\ }\textbf {\bibinfo {volume} {136}},\ \bibinfo {pages} {024315}
  (\bibinfo {year} {2012})}\BibitemShut {NoStop}%
\bibitem [{\citenamefont {Robinson}\ and\ \citenamefont
  {Knowles}(2012)}]{JCP.137.054301}%
  \BibitemOpen
  \bibfield  {author} {\bibinfo {author} {\bibfnamefont {J.~B.}\ \bibnamefont
  {Robinson}}\ and\ \bibinfo {author} {\bibfnamefont {P.~J.}\ \bibnamefont
  {Knowles}},\ }\href {\doibase 10.1063/1.4738758} {\bibfield  {journal}
  {\bibinfo  {journal} {J. Chem. Phys.}\ }\textbf {\bibinfo {volume} {137}},\
  \bibinfo {pages} {054301} (\bibinfo {year} {2012})}\BibitemShut {NoStop}%
\bibitem [{\citenamefont {White}(1992)}]{PhysRevLett.69.2863}%
  \BibitemOpen
  \bibfield  {author} {\bibinfo {author} {\bibfnamefont {S.~R.}\ \bibnamefont
  {White}},\ }\href {\doibase 10.1103/PhysRevLett.69.2863} {\bibfield
  {journal} {\bibinfo  {journal} {Phys. Rev. Lett.}\ }\textbf {\bibinfo
  {volume} {69}},\ \bibinfo {pages} {2863} (\bibinfo {year}
  {1992})}\BibitemShut {NoStop}%
\bibitem [{\citenamefont {Wouters}\ \emph {et~al.}(2012)\citenamefont
  {Wouters}, \citenamefont {Limacher}, \citenamefont {Van~Neck},\ and\
  \citenamefont {Ayers}}]{JCP.136.134110}%
  \BibitemOpen
  \bibfield  {author} {\bibinfo {author} {\bibfnamefont {S.}~\bibnamefont
  {Wouters}}, \bibinfo {author} {\bibfnamefont {P.~A.}\ \bibnamefont
  {Limacher}}, \bibinfo {author} {\bibfnamefont {D.}~\bibnamefont {Van~Neck}},
  \ and\ \bibinfo {author} {\bibfnamefont {P.~W.}\ \bibnamefont {Ayers}},\
  }\href {\doibase 10.1063/1.3700087} {\bibfield  {journal} {\bibinfo
  {journal} {J. Chem. Phys.}\ }\textbf {\bibinfo {volume} {136}},\ \bibinfo
  {pages} {134110} (\bibinfo {year} {2012})}\BibitemShut {NoStop}%
\bibitem [{\citenamefont {\"Ostlund}\ and\ \citenamefont
  {Rommer}(1995)}]{PRL.75.3537}%
  \BibitemOpen
  \bibfield  {author} {\bibinfo {author} {\bibfnamefont {S.}~\bibnamefont
  {\"Ostlund}}\ and\ \bibinfo {author} {\bibfnamefont {S.}~\bibnamefont
  {Rommer}},\ }\href {\doibase 10.1103/PhysRevLett.75.3537} {\bibfield
  {journal} {\bibinfo  {journal} {Phys. Rev. Lett.}\ }\textbf {\bibinfo
  {volume} {75}},\ \bibinfo {pages} {3537} (\bibinfo {year}
  {1995})}\BibitemShut {NoStop}%
\bibitem [{\citenamefont {White}\ and\ \citenamefont
  {Martin}(1999)}]{JCP.110.4127}%
  \BibitemOpen
  \bibfield  {author} {\bibinfo {author} {\bibfnamefont {S.~R.}\ \bibnamefont
  {White}}\ and\ \bibinfo {author} {\bibfnamefont {R.~L.}\ \bibnamefont
  {Martin}},\ }\href {\doibase 10.1063/1.478295} {\bibfield  {journal}
  {\bibinfo  {journal} {J. Chem. Phys.}\ }\textbf {\bibinfo {volume} {110}},\
  \bibinfo {pages} {4127} (\bibinfo {year} {1999})}\BibitemShut {NoStop}%
\bibitem [{\citenamefont {Mitrushenkov}\ \emph {et~al.}(2001)\citenamefont
  {Mitrushenkov}, \citenamefont {Fano}, \citenamefont {Ortolani}, \citenamefont
  {Linguerri},\ and\ \citenamefont {Palmieri}}]{JCP.115.6815}%
  \BibitemOpen
  \bibfield  {author} {\bibinfo {author} {\bibfnamefont {A.~O.}\ \bibnamefont
  {Mitrushenkov}}, \bibinfo {author} {\bibfnamefont {G.}~\bibnamefont {Fano}},
  \bibinfo {author} {\bibfnamefont {F.}~\bibnamefont {Ortolani}}, \bibinfo
  {author} {\bibfnamefont {R.}~\bibnamefont {Linguerri}}, \ and\ \bibinfo
  {author} {\bibfnamefont {P.}~\bibnamefont {Palmieri}},\ }\href {\doibase
  10.1063/1.1389475} {\bibfield  {journal} {\bibinfo  {journal} {J. Chem.
  Phys.}\ }\textbf {\bibinfo {volume} {115}},\ \bibinfo {pages} {6815}
  (\bibinfo {year} {2001})}\BibitemShut {NoStop}%
\bibitem [{\citenamefont {Chan}\ and\ \citenamefont
  {Head-Gordon}(2002)}]{JCP.116.4462}%
  \BibitemOpen
  \bibfield  {author} {\bibinfo {author} {\bibfnamefont {G.~K.-L.}\
  \bibnamefont {Chan}}\ and\ \bibinfo {author} {\bibfnamefont {M.}~\bibnamefont
  {Head-Gordon}},\ }\href {\doibase 10.1063/1.1449459} {\bibfield  {journal}
  {\bibinfo  {journal} {J. Chem. Phys.}\ }\textbf {\bibinfo {volume} {116}},\
  \bibinfo {pages} {4462} (\bibinfo {year} {2002})}\BibitemShut {NoStop}%
\bibitem [{\citenamefont {Legeza}, \citenamefont {R\"oder},\ and\ \citenamefont
  {Hess}(2003)}]{PRB.67.125114}%
  \BibitemOpen
  \bibfield  {author} {\bibinfo {author} {\bibfnamefont {O.}~\bibnamefont
  {Legeza}}, \bibinfo {author} {\bibfnamefont {J.}~\bibnamefont {R\"oder}}, \
  and\ \bibinfo {author} {\bibfnamefont {B.~A.}\ \bibnamefont {Hess}},\ }\href
  {\doibase 10.1103/PhysRevB.67.125114} {\bibfield  {journal} {\bibinfo
  {journal} {Phys. Rev. B}\ }\textbf {\bibinfo {volume} {67}},\ \bibinfo
  {pages} {125114} (\bibinfo {year} {2003})}\BibitemShut {NoStop}%
\bibitem [{\citenamefont {Moritz}, \citenamefont {Hess},\ and\ \citenamefont
  {Reiher}(2005)}]{JCP.122.024107}%
  \BibitemOpen
  \bibfield  {author} {\bibinfo {author} {\bibfnamefont {G.}~\bibnamefont
  {Moritz}}, \bibinfo {author} {\bibfnamefont {B.~A.}\ \bibnamefont {Hess}}, \
  and\ \bibinfo {author} {\bibfnamefont {M.}~\bibnamefont {Reiher}},\ }\href
  {\doibase 10.1063/1.1824891} {\bibfield  {journal} {\bibinfo  {journal} {J.
  Chem. Phys.}\ }\textbf {\bibinfo {volume} {122}},\ \bibinfo {pages} {024107}
  (\bibinfo {year} {2005})}\BibitemShut {NoStop}%
\bibitem [{\citenamefont {Zgid}\ and\ \citenamefont
  {Nooijen}(2008{\natexlab{a}})}]{JCP.128.014107}%
  \BibitemOpen
  \bibfield  {author} {\bibinfo {author} {\bibfnamefont {D.}~\bibnamefont
  {Zgid}}\ and\ \bibinfo {author} {\bibfnamefont {M.}~\bibnamefont {Nooijen}},\
  }\href {\doibase 10.1063/1.2814150} {\bibfield  {journal} {\bibinfo
  {journal} {J. Chem. Phys.}\ }\textbf {\bibinfo {volume} {128}},\ \bibinfo
  {pages} {014107} (\bibinfo {year} {2008}{\natexlab{a}})}\BibitemShut
  {NoStop}%
\bibitem [{\citenamefont {Kurashige}\ and\ \citenamefont
  {Yanai}(2009)}]{JCP.130.234114}%
  \BibitemOpen
  \bibfield  {author} {\bibinfo {author} {\bibfnamefont {Y.}~\bibnamefont
  {Kurashige}}\ and\ \bibinfo {author} {\bibfnamefont {T.}~\bibnamefont
  {Yanai}},\ }\href {\doibase 10.1063/1.3152576} {\bibfield  {journal}
  {\bibinfo  {journal} {J. Chem. Phys.}\ }\textbf {\bibinfo {volume} {130}},\
  \bibinfo {pages} {234114} (\bibinfo {year} {2009})}\BibitemShut {NoStop}%
\bibitem [{\citenamefont {Luo}, \citenamefont {Qin},\ and\ \citenamefont
  {Xiang}(2010)}]{PRB.81.235129}%
  \BibitemOpen
  \bibfield  {author} {\bibinfo {author} {\bibfnamefont {H.-G.}\ \bibnamefont
  {Luo}}, \bibinfo {author} {\bibfnamefont {M.-P.}\ \bibnamefont {Qin}}, \ and\
  \bibinfo {author} {\bibfnamefont {T.}~\bibnamefont {Xiang}},\ }\href
  {\doibase 10.1103/PhysRevB.81.235129} {\bibfield  {journal} {\bibinfo
  {journal} {Phys. Rev. B}\ }\textbf {\bibinfo {volume} {81}},\ \bibinfo
  {pages} {235129} (\bibinfo {year} {2010})}\BibitemShut {NoStop}%
\bibitem [{\citenamefont {Wouters}\ \emph
  {et~al.}(2014{\natexlab{a}})\citenamefont {Wouters}, \citenamefont
  {Poelmans}, \citenamefont {Ayers},\ and\ \citenamefont {{Van
  Neck}}}]{CPC.185.1501}%
  \BibitemOpen
  \bibfield  {author} {\bibinfo {author} {\bibfnamefont {S.}~\bibnamefont
  {Wouters}}, \bibinfo {author} {\bibfnamefont {W.}~\bibnamefont {Poelmans}},
  \bibinfo {author} {\bibfnamefont {P.~W.}\ \bibnamefont {Ayers}}, \ and\
  \bibinfo {author} {\bibfnamefont {D.}~\bibnamefont {{Van Neck}}},\ }\href
  {\doibase 10.1016/j.cpc.2014.01.019} {\bibfield  {journal} {\bibinfo
  {journal} {Comput. Phys. Commun.}\ }\textbf {\bibinfo {volume} {185}},\
  \bibinfo {pages} {1501} (\bibinfo {year} {2014}{\natexlab{a}})}\BibitemShut
  {NoStop}%
\bibitem [{\citenamefont {Keller}\ \emph
  {et~al.}(2015{\natexlab{a}})\citenamefont {Keller}, \citenamefont {Dolfi},
  \citenamefont {Troyer},\ and\ \citenamefont {Reiher}}]{JCP.143.244118}%
  \BibitemOpen
  \bibfield  {author} {\bibinfo {author} {\bibfnamefont {S.}~\bibnamefont
  {Keller}}, \bibinfo {author} {\bibfnamefont {M.}~\bibnamefont {Dolfi}},
  \bibinfo {author} {\bibfnamefont {M.}~\bibnamefont {Troyer}}, \ and\ \bibinfo
  {author} {\bibfnamefont {M.}~\bibnamefont {Reiher}},\ }\href {\doibase
  10.1063/1.4939000} {\bibfield  {journal} {\bibinfo  {journal} {J. Chem.
  Phys.}\ }\textbf {\bibinfo {volume} {143}},\ \bibinfo {pages} {244118}
  (\bibinfo {year} {2015}{\natexlab{a}})}\BibitemShut {NoStop}%
\bibitem [{\citenamefont {Zgid}\ and\ \citenamefont
  {Nooijen}(2008{\natexlab{b}})}]{JCP.128.144116}%
  \BibitemOpen
  \bibfield  {author} {\bibinfo {author} {\bibfnamefont {D.}~\bibnamefont
  {Zgid}}\ and\ \bibinfo {author} {\bibfnamefont {M.}~\bibnamefont {Nooijen}},\
  }\href {\doibase 10.1063/1.2883981} {\bibfield  {journal} {\bibinfo
  {journal} {J. Chem. Phys.}\ }\textbf {\bibinfo {volume} {128}},\ \bibinfo
  {pages} {144116} (\bibinfo {year} {2008}{\natexlab{b}})}\BibitemShut
  {NoStop}%
\bibitem [{\citenamefont {Ghosh}\ \emph {et~al.}(2008)\citenamefont {Ghosh},
  \citenamefont {Hachmann}, \citenamefont {Yanai},\ and\ \citenamefont
  {Chan}}]{JCP.128.144117}%
  \BibitemOpen
  \bibfield  {author} {\bibinfo {author} {\bibfnamefont {D.}~\bibnamefont
  {Ghosh}}, \bibinfo {author} {\bibfnamefont {J.}~\bibnamefont {Hachmann}},
  \bibinfo {author} {\bibfnamefont {T.}~\bibnamefont {Yanai}}, \ and\ \bibinfo
  {author} {\bibfnamefont {G.~K.-L.}\ \bibnamefont {Chan}},\ }\href {\doibase
  10.1063/1.2883976} {\bibfield  {journal} {\bibinfo  {journal} {J. Chem.
  Phys.}\ }\textbf {\bibinfo {volume} {128}},\ \bibinfo {pages} {144117}
  (\bibinfo {year} {2008})}\BibitemShut {NoStop}%
\bibitem [{\citenamefont {Yanai}\ \emph {et~al.}(2009)\citenamefont {Yanai},
  \citenamefont {Kurashige}, \citenamefont {Ghosh},\ and\ \citenamefont
  {Chan}}]{IJQC.109.2178}%
  \BibitemOpen
  \bibfield  {author} {\bibinfo {author} {\bibfnamefont {T.}~\bibnamefont
  {Yanai}}, \bibinfo {author} {\bibfnamefont {Y.}~\bibnamefont {Kurashige}},
  \bibinfo {author} {\bibfnamefont {D.}~\bibnamefont {Ghosh}}, \ and\ \bibinfo
  {author} {\bibfnamefont {G.~K.-L.}\ \bibnamefont {Chan}},\ }\href {\doibase
  10.1002/qua.22099} {\bibfield  {journal} {\bibinfo  {journal} {Int. J. Quant.
  Chem.}\ }\textbf {\bibinfo {volume} {109}},\ \bibinfo {pages} {2178}
  (\bibinfo {year} {2009})}\BibitemShut {NoStop}%
\bibitem [{\citenamefont {Wouters}\ \emph
  {et~al.}(2014{\natexlab{b}})\citenamefont {Wouters}, \citenamefont
  {Bogaerts}, \citenamefont {Van Der~Voort}, \citenamefont {Van~Speybroeck},\
  and\ \citenamefont {Van~Neck}}]{JCP.140.241103}%
  \BibitemOpen
  \bibfield  {author} {\bibinfo {author} {\bibfnamefont {S.}~\bibnamefont
  {Wouters}}, \bibinfo {author} {\bibfnamefont {T.}~\bibnamefont {Bogaerts}},
  \bibinfo {author} {\bibfnamefont {P.}~\bibnamefont {Van Der~Voort}}, \bibinfo
  {author} {\bibfnamefont {V.}~\bibnamefont {Van~Speybroeck}}, \ and\ \bibinfo
  {author} {\bibfnamefont {D.}~\bibnamefont {Van~Neck}},\ }\href {\doibase
  10.1063/1.4885815} {\bibfield  {journal} {\bibinfo  {journal} {J. Chem.
  Phys.}\ }\textbf {\bibinfo {volume} {140}},\ \bibinfo {pages} {241103}
  (\bibinfo {year} {2014}{\natexlab{b}})}\BibitemShut {NoStop}%
\bibitem [{\citenamefont {Yanai}\ \emph {et~al.}(2010)\citenamefont {Yanai},
  \citenamefont {Kurashige}, \citenamefont {Neuscamman},\ and\ \citenamefont
  {Chan}}]{JCP.132.024105}%
  \BibitemOpen
  \bibfield  {author} {\bibinfo {author} {\bibfnamefont {T.}~\bibnamefont
  {Yanai}}, \bibinfo {author} {\bibfnamefont {Y.}~\bibnamefont {Kurashige}},
  \bibinfo {author} {\bibfnamefont {E.}~\bibnamefont {Neuscamman}}, \ and\
  \bibinfo {author} {\bibfnamefont {G.~K.-L.}\ \bibnamefont {Chan}},\ }\href
  {\doibase 10.1063/1.3275806} {\bibfield  {journal} {\bibinfo  {journal} {J.
  Chem. Phys.}\ }\textbf {\bibinfo {volume} {132}},\ \bibinfo {pages} {024105}
  (\bibinfo {year} {2010})}\BibitemShut {NoStop}%
\bibitem [{\citenamefont {Yanai}\ \emph {et~al.}(2012)\citenamefont {Yanai},
  \citenamefont {Kurashige}, \citenamefont {Neuscamman},\ and\ \citenamefont
  {Chan}}]{PCCP.14.7809}%
  \BibitemOpen
  \bibfield  {author} {\bibinfo {author} {\bibfnamefont {T.}~\bibnamefont
  {Yanai}}, \bibinfo {author} {\bibfnamefont {Y.}~\bibnamefont {Kurashige}},
  \bibinfo {author} {\bibfnamefont {E.}~\bibnamefont {Neuscamman}}, \ and\
  \bibinfo {author} {\bibfnamefont {G.~K.-L.}\ \bibnamefont {Chan}},\ }\href
  {\doibase 10.1039/C2CP23767A} {\bibfield  {journal} {\bibinfo  {journal}
  {Phys. Chem. Chem. Phys.}\ }\textbf {\bibinfo {volume} {14}},\ \bibinfo
  {pages} {7809} (\bibinfo {year} {2012})}\BibitemShut {NoStop}%
\bibitem [{\citenamefont {Kurashige}\ and\ \citenamefont
  {Yanai}(2011)}]{JCP.135.094104}%
  \BibitemOpen
  \bibfield  {author} {\bibinfo {author} {\bibfnamefont {Y.}~\bibnamefont
  {Kurashige}}\ and\ \bibinfo {author} {\bibfnamefont {T.}~\bibnamefont
  {Yanai}},\ }\href {\doibase 10.1063/1.3629454} {\bibfield  {journal}
  {\bibinfo  {journal} {J. Chem. Phys.}\ }\textbf {\bibinfo {volume} {135}},\
  \bibinfo {pages} {094104} (\bibinfo {year} {2011})}\BibitemShut {NoStop}%
\bibitem [{\citenamefont {Kurashige}\ \emph {et~al.}(2014)\citenamefont
  {Kurashige}, \citenamefont {Chalupsk\'y}, \citenamefont {Lan},\ and\
  \citenamefont {Yanai}}]{JCP.141.174111}%
  \BibitemOpen
  \bibfield  {author} {\bibinfo {author} {\bibfnamefont {Y.}~\bibnamefont
  {Kurashige}}, \bibinfo {author} {\bibfnamefont {J.}~\bibnamefont
  {Chalupsk\'y}}, \bibinfo {author} {\bibfnamefont {T.~N.}\ \bibnamefont
  {Lan}}, \ and\ \bibinfo {author} {\bibfnamefont {T.}~\bibnamefont {Yanai}},\
  }\href {\doibase 10.1063/1.4900878} {\bibfield  {journal} {\bibinfo
  {journal} {J. Chem. Phys.}\ }\textbf {\bibinfo {volume} {141}},\ \bibinfo
  {pages} {174111} (\bibinfo {year} {2014})}\BibitemShut {NoStop}%
\bibitem [{\citenamefont {Guo}\ \emph {et~al.}(2016)\citenamefont {Guo},
  \citenamefont {Watson}, \citenamefont {Hu}, \citenamefont {Sun},\ and\
  \citenamefont {Chan}}]{JCTC.12.1583}%
  \BibitemOpen
  \bibfield  {author} {\bibinfo {author} {\bibfnamefont {S.}~\bibnamefont
  {Guo}}, \bibinfo {author} {\bibfnamefont {M.~A.}\ \bibnamefont {Watson}},
  \bibinfo {author} {\bibfnamefont {W.}~\bibnamefont {Hu}}, \bibinfo {author}
  {\bibfnamefont {Q.}~\bibnamefont {Sun}}, \ and\ \bibinfo {author}
  {\bibfnamefont {G.~K.-L.}\ \bibnamefont {Chan}},\ }\href {\doibase
  10.1021/acs.jctc.5b01225} {\bibfield  {journal} {\bibinfo  {journal} {J.
  Chem. Theory Comput.}\ }\textbf {\bibinfo {volume} {12}},\ \bibinfo {pages}
  {1583} (\bibinfo {year} {2016})}\BibitemShut {NoStop}%
\bibitem [{\citenamefont {Saitow}, \citenamefont {Kurashige},\ and\
  \citenamefont {Yanai}(2013)}]{JCP.139.044118}%
  \BibitemOpen
  \bibfield  {author} {\bibinfo {author} {\bibfnamefont {M.}~\bibnamefont
  {Saitow}}, \bibinfo {author} {\bibfnamefont {Y.}~\bibnamefont {Kurashige}}, \
  and\ \bibinfo {author} {\bibfnamefont {T.}~\bibnamefont {Yanai}},\ }\href
  {\doibase 10.1063/1.4816627} {\bibfield  {journal} {\bibinfo  {journal} {J.
  Chem. Phys.}\ }\textbf {\bibinfo {volume} {139}},\ \bibinfo {pages} {044118}
  (\bibinfo {year} {2013})}\BibitemShut {NoStop}%
\bibitem [{\citenamefont {Saitow}, \citenamefont {Kurashige},\ and\
  \citenamefont {Yanai}(2015)}]{JCTC.11.5120}%
  \BibitemOpen
  \bibfield  {author} {\bibinfo {author} {\bibfnamefont {M.}~\bibnamefont
  {Saitow}}, \bibinfo {author} {\bibfnamefont {Y.}~\bibnamefont {Kurashige}}, \
  and\ \bibinfo {author} {\bibfnamefont {T.}~\bibnamefont {Yanai}},\ }\href
  {\doibase 10.1021/acs.jctc.5b00270} {\bibfield  {journal} {\bibinfo
  {journal} {J. Chem. Theory Comput.}\ }\textbf {\bibinfo {volume} {11}},\
  \bibinfo {pages} {5120} (\bibinfo {year} {2015})}\BibitemShut {NoStop}%
\bibitem [{\citenamefont {Sharma}\ and\ \citenamefont
  {Chan}(2014)}]{JCP.141.111101}%
  \BibitemOpen
  \bibfield  {author} {\bibinfo {author} {\bibfnamefont {S.}~\bibnamefont
  {Sharma}}\ and\ \bibinfo {author} {\bibfnamefont {G.~K.-L.}\ \bibnamefont
  {Chan}},\ }\href {\doibase 10.1063/1.4895977} {\bibfield  {journal} {\bibinfo
   {journal} {J. Chem. Phys.}\ }\textbf {\bibinfo {volume} {141}},\ \bibinfo
  {pages} {111101} (\bibinfo {year} {2014})}\BibitemShut {NoStop}%
\bibitem [{\citenamefont {Sharma}\ and\ \citenamefont
  {Alavi}(2015)}]{JCP.143.102815}%
  \BibitemOpen
  \bibfield  {author} {\bibinfo {author} {\bibfnamefont {S.}~\bibnamefont
  {Sharma}}\ and\ \bibinfo {author} {\bibfnamefont {A.}~\bibnamefont {Alavi}},\
  }\href {\doibase 10.1063/1.4928643} {\bibfield  {journal} {\bibinfo
  {journal} {J. Chem. Phys.}\ }\textbf {\bibinfo {volume} {143}},\ \bibinfo
  {eid} {102815} (\bibinfo {year} {2015})}\BibitemShut {NoStop}%
\bibitem [{\citenamefont {Sharma}, \citenamefont {Jeanmairet},\ and\
  \citenamefont {Alavi}(2016)}]{JCP.144.034103}%
  \BibitemOpen
  \bibfield  {author} {\bibinfo {author} {\bibfnamefont {S.}~\bibnamefont
  {Sharma}}, \bibinfo {author} {\bibfnamefont {G.}~\bibnamefont {Jeanmairet}},
  \ and\ \bibinfo {author} {\bibfnamefont {A.}~\bibnamefont {Alavi}},\ }\href
  {\doibase 10.1063/1.4939752} {\bibfield  {journal} {\bibinfo  {journal} {J.
  Chem. Phys.}\ }\textbf {\bibinfo {volume} {144}},\ \bibinfo {pages} {034103}
  (\bibinfo {year} {2016})}\BibitemShut {NoStop}%
\bibitem [{\citenamefont {Kurashige}, \citenamefont {Chan},\ and\ \citenamefont
  {Yanai}(2013)}]{NC.5.660}%
  \BibitemOpen
  \bibfield  {author} {\bibinfo {author} {\bibfnamefont {Y.}~\bibnamefont
  {Kurashige}}, \bibinfo {author} {\bibfnamefont {G.~K.-L.}\ \bibnamefont
  {Chan}}, \ and\ \bibinfo {author} {\bibfnamefont {T.}~\bibnamefont {Yanai}},\
  }\href {\doibase 10.1038/nchem.1677} {\bibfield  {journal} {\bibinfo
  {journal} {Nat. Chem.}\ }\textbf {\bibinfo {volume} {5}},\ \bibinfo {pages}
  {660} (\bibinfo {year} {2013})}\BibitemShut {NoStop}%
\bibitem [{\citenamefont {Sharma}\ \emph
  {et~al.}(2014{\natexlab{a}})\citenamefont {Sharma}, \citenamefont
  {Sivalingam}, \citenamefont {Neese},\ and\ \citenamefont {Chan}}]{NC.6.927}%
  \BibitemOpen
  \bibfield  {author} {\bibinfo {author} {\bibfnamefont {S.}~\bibnamefont
  {Sharma}}, \bibinfo {author} {\bibfnamefont {K.}~\bibnamefont {Sivalingam}},
  \bibinfo {author} {\bibfnamefont {F.}~\bibnamefont {Neese}}, \ and\ \bibinfo
  {author} {\bibfnamefont {G.~K.-L.}\ \bibnamefont {Chan}},\ }\href {\doibase
  10.1038/nchem.2041} {\bibfield  {journal} {\bibinfo  {journal} {Nat. Chem.}\
  }\textbf {\bibinfo {volume} {6}},\ \bibinfo {pages} {927} (\bibinfo {year}
  {2014}{\natexlab{a}})}\BibitemShut {NoStop}%
\bibitem [{\citenamefont {Chalupsk\'y}\ \emph {et~al.}(2014)\citenamefont
  {Chalupsk\'y}, \citenamefont {Rokob}, \citenamefont {Kurashige},
  \citenamefont {Yanai}, \citenamefont {Solomon}, \citenamefont {Rul\'isek},\
  and\ \citenamefont {Srnec}}]{JACS.136.15977}%
  \BibitemOpen
  \bibfield  {author} {\bibinfo {author} {\bibfnamefont {J.}~\bibnamefont
  {Chalupsk\'y}}, \bibinfo {author} {\bibfnamefont {T.~A.}\ \bibnamefont
  {Rokob}}, \bibinfo {author} {\bibfnamefont {Y.}~\bibnamefont {Kurashige}},
  \bibinfo {author} {\bibfnamefont {T.}~\bibnamefont {Yanai}}, \bibinfo
  {author} {\bibfnamefont {E.~I.}\ \bibnamefont {Solomon}}, \bibinfo {author}
  {\bibfnamefont {L.}~\bibnamefont {Rul\'isek}}, \ and\ \bibinfo {author}
  {\bibfnamefont {M.}~\bibnamefont {Srnec}},\ }\href {\doibase
  10.1021/ja506934k} {\bibfield  {journal} {\bibinfo  {journal} {J. Am. Chem.
  Soc.}\ }\textbf {\bibinfo {volume} {136}},\ \bibinfo {pages} {15977}
  (\bibinfo {year} {2014})}\BibitemShut {NoStop}%
\bibitem [{\citenamefont {Hastings}(2007)}]{JSMTE.2007.P08024}%
  \BibitemOpen
  \bibfield  {author} {\bibinfo {author} {\bibfnamefont {M.~B.}\ \bibnamefont
  {Hastings}},\ }\href {\doibase 10.1088/1742-5468/2007/08/P08024} {\bibfield
  {journal} {\bibinfo  {journal} {J. Stat. Mech.: Theory Exp.}\ }\textbf
  {\bibinfo {volume} {2007}},\ \bibinfo {pages} {P08024} (\bibinfo {year}
  {2007})}\BibitemShut {NoStop}%
\bibitem [{\citenamefont {Wouters}\ and\ \citenamefont
  {Neck}(2014)}]{EPJD.68.272}%
  \BibitemOpen
  \bibfield  {author} {\bibinfo {author} {\bibfnamefont {S.}~\bibnamefont
  {Wouters}}\ and\ \bibinfo {author} {\bibfnamefont {D.}~\bibnamefont {Neck}},\
  }\href {\doibase 10.1140/epjd/e2014-50500-1} {\bibfield  {journal} {\bibinfo
  {journal} {Eur. Phys. J. D}\ }\textbf {\bibinfo {volume} {68}},\ \bibinfo
  {pages} {272} (\bibinfo {year} {2014})}\BibitemShut {NoStop}%
\bibitem [{\citenamefont {Sierra}\ and\ \citenamefont
  {Nishino}(1997)}]{NPB.495.505}%
  \BibitemOpen
  \bibfield  {author} {\bibinfo {author} {\bibfnamefont {G.}~\bibnamefont
  {Sierra}}\ and\ \bibinfo {author} {\bibfnamefont {T.}~\bibnamefont
  {Nishino}},\ }\href {\doibase 10.1016/S0550-3213(97)00217-4} {\bibfield
  {journal} {\bibinfo  {journal} {Nucl. Phys. B}\ }\textbf {\bibinfo {volume}
  {495}},\ \bibinfo {pages} {505} (\bibinfo {year} {1997})}\BibitemShut
  {NoStop}%
\bibitem [{\citenamefont {McCulloch}\ and\ \citenamefont
  {Gul\'acsi}(2002)}]{EPL.57.852}%
  \BibitemOpen
  \bibfield  {author} {\bibinfo {author} {\bibfnamefont {I.~P.}\ \bibnamefont
  {McCulloch}}\ and\ \bibinfo {author} {\bibfnamefont {M.}~\bibnamefont
  {Gul\'acsi}},\ }\href {\doibase 10.1209/epl/i2002-00393-0} {\bibfield
  {journal} {\bibinfo  {journal} {EPL (Europhysics Letters)}\ }\textbf
  {\bibinfo {volume} {57}},\ \bibinfo {pages} {852} (\bibinfo {year}
  {2002})}\BibitemShut {NoStop}%
\bibitem [{\citenamefont {Pittel}\ and\ \citenamefont
  {Sandulescu}(2006)}]{PRC.73.014301}%
  \BibitemOpen
  \bibfield  {author} {\bibinfo {author} {\bibfnamefont {S.}~\bibnamefont
  {Pittel}}\ and\ \bibinfo {author} {\bibfnamefont {N.}~\bibnamefont
  {Sandulescu}},\ }\href {\doibase 10.1103/PhysRevC.73.014301} {\bibfield
  {journal} {\bibinfo  {journal} {Phys. Rev. C}\ }\textbf {\bibinfo {volume}
  {73}},\ \bibinfo {pages} {014301} (\bibinfo {year} {2006})}\BibitemShut
  {NoStop}%
\bibitem [{\citenamefont {T\'oth}\ \emph {et~al.}(2008)\citenamefont {T\'oth},
  \citenamefont {Moca}, \citenamefont {Legeza},\ and\ \citenamefont
  {Zar\'and}}]{PRB.78.245109}%
  \BibitemOpen
  \bibfield  {author} {\bibinfo {author} {\bibfnamefont {A.~I.}\ \bibnamefont
  {T\'oth}}, \bibinfo {author} {\bibfnamefont {C.~P.}\ \bibnamefont {Moca}},
  \bibinfo {author} {\bibfnamefont {O.}~\bibnamefont {Legeza}}, \ and\ \bibinfo
  {author} {\bibfnamefont {G.}~\bibnamefont {Zar\'and}},\ }\href {\doibase
  10.1103/PhysRevB.78.245109} {\bibfield  {journal} {\bibinfo  {journal} {Phys.
  Rev. B}\ }\textbf {\bibinfo {volume} {78}},\ \bibinfo {pages} {245109}
  (\bibinfo {year} {2008})}\BibitemShut {NoStop}%
\bibitem [{\citenamefont {Sharma}\ and\ \citenamefont
  {Chan}(2012)}]{JCP.136.124121}%
  \BibitemOpen
  \bibfield  {author} {\bibinfo {author} {\bibfnamefont {S.}~\bibnamefont
  {Sharma}}\ and\ \bibinfo {author} {\bibfnamefont {G.~K.-L.}\ \bibnamefont
  {Chan}},\ }\href {\doibase 10.1063/1.3695642} {\bibfield  {journal} {\bibinfo
   {journal} {J. Chem. Phys.}\ }\textbf {\bibinfo {volume} {136}},\ \bibinfo
  {pages} {124121} (\bibinfo {year} {2012})}\BibitemShut {NoStop}%
\bibitem [{\citenamefont {Keller}\ and\ \citenamefont
  {Reiher}(2016)}]{JCP.144.134101}%
  \BibitemOpen
  \bibfield  {author} {\bibinfo {author} {\bibfnamefont {S.}~\bibnamefont
  {Keller}}\ and\ \bibinfo {author} {\bibfnamefont {M.}~\bibnamefont
  {Reiher}},\ }\href {\doibase 10.1063/1.4944921} {\bibfield  {journal}
  {\bibinfo  {journal} {J. Chem. Phys.}\ }\textbf {\bibinfo {volume} {144}},\
  \bibinfo {pages} {134101} (\bibinfo {year} {2016})}\BibitemShut {NoStop}%
\bibitem [{\citenamefont {Sharma}\ \emph
  {et~al.}(2014{\natexlab{b}})\citenamefont {Sharma}, \citenamefont {Yanai},
  \citenamefont {Booth}, \citenamefont {Umrigar},\ and\ \citenamefont
  {Chan}}]{JCP.140.104112}%
  \BibitemOpen
  \bibfield  {author} {\bibinfo {author} {\bibfnamefont {S.}~\bibnamefont
  {Sharma}}, \bibinfo {author} {\bibfnamefont {T.}~\bibnamefont {Yanai}},
  \bibinfo {author} {\bibfnamefont {G.~H.}\ \bibnamefont {Booth}}, \bibinfo
  {author} {\bibfnamefont {C.~J.}\ \bibnamefont {Umrigar}}, \ and\ \bibinfo
  {author} {\bibfnamefont {G.~K.-L.}\ \bibnamefont {Chan}},\ }\href {\doibase
  10.1063/1.4867383} {\bibfield  {journal} {\bibinfo  {journal} {J. Chem.
  Phys.}\ }\textbf {\bibinfo {volume} {140}},\ \bibinfo {pages} {104112}
  (\bibinfo {year} {2014}{\natexlab{b}})}\BibitemShut {NoStop}%
\bibitem [{\citenamefont {Sharma}(2015)}]{JCP.142.024107}%
  \BibitemOpen
  \bibfield  {author} {\bibinfo {author} {\bibfnamefont {S.}~\bibnamefont
  {Sharma}},\ }\href {\doibase 10.1063/1.4905237} {\bibfield  {journal}
  {\bibinfo  {journal} {J. Chem. Phys.}\ }\textbf {\bibinfo {volume} {142}},\
  \bibinfo {pages} {024107} (\bibinfo {year} {2015})}\BibitemShut {NoStop}%
\bibitem [{\citenamefont {Shepard}, \citenamefont {Gidofalvi},\ and\
  \citenamefont {Brozell}(2014)}]{JCP.141.064105}%
  \BibitemOpen
  \bibfield  {author} {\bibinfo {author} {\bibfnamefont {R.}~\bibnamefont
  {Shepard}}, \bibinfo {author} {\bibfnamefont {G.}~\bibnamefont {Gidofalvi}},
  \ and\ \bibinfo {author} {\bibfnamefont {S.~R.}\ \bibnamefont {Brozell}},\
  }\href {\doibase 10.1063/1.4890734} {\bibfield  {journal} {\bibinfo
  {journal} {J. Chem. Phys.}\ }\textbf {\bibinfo {volume} {141}},\ \bibinfo
  {pages} {064105} (\bibinfo {year} {2014})}\BibitemShut {NoStop}%
\bibitem [{\citenamefont {Sharma}\ and\ \citenamefont
  {Chan}(2016)}]{block-1.1-alpha}%
  \BibitemOpen
  \bibfield  {author} {\bibinfo {author} {\bibfnamefont {S.}~\bibnamefont
  {Sharma}}\ and\ \bibinfo {author} {\bibfnamefont {G.~K.-L.}\ \bibnamefont
  {Chan}},\ }\href@noop {} {} (\bibinfo {year} {2016}),\ \bibinfo {note}
  {\textsc{block} version 1.1-alpha,
  \url{https://github.com/sanshar/block/releases/tag/v1.1-alpha}}\BibitemShut
  {NoStop}%
\bibitem [{\citenamefont {Chan}\ and\ \citenamefont
  {Head-Gordon}(2003)}]{JCP.118.8551}%
  \BibitemOpen
  \bibfield  {author} {\bibinfo {author} {\bibfnamefont {G.~K.-L.}\
  \bibnamefont {Chan}}\ and\ \bibinfo {author} {\bibfnamefont {M.}~\bibnamefont
  {Head-Gordon}},\ }\href {\doibase 10.1063/1.1574318} {\bibfield  {journal}
  {\bibinfo  {journal} {J. Chem. Phys.}\ }\textbf {\bibinfo {volume} {118}},\
  \bibinfo {pages} {8551} (\bibinfo {year} {2003})}\BibitemShut {NoStop}%
\bibitem [{mpi()}]{mpi_reference}%
  \BibitemOpen
  \href@noop {} {}\bibinfo {note} {Message Passing Interface,
  \url{http://www.mpi-forum.org/}}\BibitemShut {NoStop}%
\bibitem [{\citenamefont {Chan}(2004)}]{JCP.120.3172}%
  \BibitemOpen
  \bibfield  {author} {\bibinfo {author} {\bibfnamefont {G.~K.-L.}\
  \bibnamefont {Chan}},\ }\href {\doibase 10.1063/1.1638734} {\bibfield
  {journal} {\bibinfo  {journal} {J. Chem. Phys.}\ }\textbf {\bibinfo {volume}
  {120}},\ \bibinfo {pages} {3172} (\bibinfo {year} {2004})}\BibitemShut
  {NoStop}%
\bibitem [{omp()}]{omp_reference}%
  \BibitemOpen
  \href@noop {} {}\bibinfo {note} {Open Multi-Processing API,
  \url{http://openmp.org/wp/}}\BibitemShut {NoStop}%
\bibitem [{\citenamefont {Stoudenmire}\ and\ \citenamefont
  {White}(2013)}]{PRB.87.155137}%
  \BibitemOpen
  \bibfield  {author} {\bibinfo {author} {\bibfnamefont {E.~M.}\ \bibnamefont
  {Stoudenmire}}\ and\ \bibinfo {author} {\bibfnamefont {S.~R.}\ \bibnamefont
  {White}},\ }\href {\doibase 10.1103/PhysRevB.87.155137} {\bibfield  {journal}
  {\bibinfo  {journal} {Phys. Rev. B}\ }\textbf {\bibinfo {volume} {87}},\
  \bibinfo {pages} {155137} (\bibinfo {year} {2013})}\BibitemShut {NoStop}%
\bibitem [{\citenamefont {Zgid}\ and\ \citenamefont
  {Nooijen}(2008{\natexlab{c}})}]{JCP.128.144115}%
  \BibitemOpen
  \bibfield  {author} {\bibinfo {author} {\bibfnamefont {D.}~\bibnamefont
  {Zgid}}\ and\ \bibinfo {author} {\bibfnamefont {M.}~\bibnamefont {Nooijen}},\
  }\href {\doibase 10.1063/1.2883980} {\bibfield  {journal} {\bibinfo
  {journal} {J. Chem. Phys.}\ }\textbf {\bibinfo {volume} {128}},\ \bibinfo
  {pages} {144115} (\bibinfo {year} {2008}{\natexlab{c}})}\BibitemShut
  {NoStop}%
\bibitem [{\citenamefont {Guo}\ and\ \citenamefont
  {Chan}(2016)}]{perturber-page}%
  \BibitemOpen
  \bibfield  {author} {\bibinfo {author} {\bibfnamefont {S.}~\bibnamefont
  {Guo}}\ and\ \bibinfo {author} {\bibfnamefont {G.~K.-L.}\ \bibnamefont
  {Chan}},\ }\href@noop {} {} (\bibinfo {year} {2016}),\ \bibinfo {note}
  {\url{http://sanshar.github.io/Block/examples-with-pyscf.html\#dmrg-nevpt2}}\BibitemShut
  {NoStop}%
\bibitem [{\citenamefont {Siegbahn}\ \emph {et~al.}(1981)\citenamefont
  {Siegbahn}, \citenamefont {Alml\"of}, \citenamefont {Heiberg},\ and\
  \citenamefont {Roos}}]{JCP.74.2384}%
  \BibitemOpen
  \bibfield  {author} {\bibinfo {author} {\bibfnamefont {P.~E.~M.}\
  \bibnamefont {Siegbahn}}, \bibinfo {author} {\bibfnamefont {J.}~\bibnamefont
  {Alml\"of}}, \bibinfo {author} {\bibfnamefont {A.}~\bibnamefont {Heiberg}}, \
  and\ \bibinfo {author} {\bibfnamefont {B.~O.}\ \bibnamefont {Roos}},\ }\href
  {\doibase 10.1063/1.441359} {\bibfield  {journal} {\bibinfo  {journal} {J.
  Chem. Phys.}\ }\textbf {\bibinfo {volume} {74}},\ \bibinfo {pages} {2384}
  (\bibinfo {year} {1981})}\BibitemShut {NoStop}%
\bibitem [{\citenamefont {Banerjee}\ \emph {et~al.}(1985)\citenamefont
  {Banerjee}, \citenamefont {Adams}, \citenamefont {Simons},\ and\
  \citenamefont {Shepard}}]{JPC.89.52}%
  \BibitemOpen
  \bibfield  {author} {\bibinfo {author} {\bibfnamefont {A.}~\bibnamefont
  {Banerjee}}, \bibinfo {author} {\bibfnamefont {N.}~\bibnamefont {Adams}},
  \bibinfo {author} {\bibfnamefont {J.}~\bibnamefont {Simons}}, \ and\ \bibinfo
  {author} {\bibfnamefont {R.}~\bibnamefont {Shepard}},\ }\href {\doibase
  10.1021/j100247a015} {\bibfield  {journal} {\bibinfo  {journal} {J. Phys.
  Chem.}\ }\textbf {\bibinfo {volume} {89}},\ \bibinfo {pages} {52} (\bibinfo
  {year} {1985})}\BibitemShut {NoStop}%
\bibitem [{\citenamefont {Keller}\ \emph
  {et~al.}(2015{\natexlab{b}})\citenamefont {Keller}, \citenamefont
  {Boguslawski}, \citenamefont {Janowski}, \citenamefont {Reiher},\ and\
  \citenamefont {Pulay}}]{JCP.142.244104}%
  \BibitemOpen
  \bibfield  {author} {\bibinfo {author} {\bibfnamefont {S.}~\bibnamefont
  {Keller}}, \bibinfo {author} {\bibfnamefont {K.}~\bibnamefont {Boguslawski}},
  \bibinfo {author} {\bibfnamefont {T.}~\bibnamefont {Janowski}}, \bibinfo
  {author} {\bibfnamefont {M.}~\bibnamefont {Reiher}}, \ and\ \bibinfo {author}
  {\bibfnamefont {P.}~\bibnamefont {Pulay}},\ }\href {\doibase
  10.1063/1.4922352} {\bibfield  {journal} {\bibinfo  {journal} {J. Chem.
  Phys.}\ }\textbf {\bibinfo {volume} {142}},\ \bibinfo {pages} {244104}
  (\bibinfo {year} {2015}{\natexlab{b}})}\BibitemShut {NoStop}%
\bibitem [{\citenamefont {Stein}\ and\ \citenamefont
  {Reiher}(2016)}]{JCTC.12.1760}%
  \BibitemOpen
  \bibfield  {author} {\bibinfo {author} {\bibfnamefont {C.~J.}\ \bibnamefont
  {Stein}}\ and\ \bibinfo {author} {\bibfnamefont {M.}~\bibnamefont {Reiher}},\
  }\href {\doibase 10.1021/acs.jctc.6b00156} {\bibfield  {journal} {\bibinfo
  {journal} {J. Chem. Theory Comput.}\ }\textbf {\bibinfo {volume} {12}},\
  \bibinfo {pages} {1760} (\bibinfo {year} {2016})}\BibitemShut {NoStop}%
\bibitem [{\citenamefont {Overy}\ \emph {et~al.}(2014)\citenamefont {Overy},
  \citenamefont {Booth}, \citenamefont {Blunt}, \citenamefont {Shepherd},
  \citenamefont {Cleland},\ and\ \citenamefont {Alavi}}]{JCP.141.244117}%
  \BibitemOpen
  \bibfield  {author} {\bibinfo {author} {\bibfnamefont {C.}~\bibnamefont
  {Overy}}, \bibinfo {author} {\bibfnamefont {G.~H.}\ \bibnamefont {Booth}},
  \bibinfo {author} {\bibfnamefont {N.~S.}\ \bibnamefont {Blunt}}, \bibinfo
  {author} {\bibfnamefont {J.~J.}\ \bibnamefont {Shepherd}}, \bibinfo {author}
  {\bibfnamefont {D.}~\bibnamefont {Cleland}}, \ and\ \bibinfo {author}
  {\bibfnamefont {A.}~\bibnamefont {Alavi}},\ }\href {\doibase
  10.1063/1.4904313} {\bibfield  {journal} {\bibinfo  {journal} {J. Chem.
  Phys.}\ }\textbf {\bibinfo {volume} {141}},\ \bibinfo {pages} {244117}
  (\bibinfo {year} {2014})}\BibitemShut {NoStop}%
\bibitem [{\citenamefont {Manni}, \citenamefont {Smart},\ and\ \citenamefont
  {Alavi}(2016)}]{JCTC.12.1245}%
  \BibitemOpen
  \bibfield  {author} {\bibinfo {author} {\bibfnamefont {G.~L.}\ \bibnamefont
  {Manni}}, \bibinfo {author} {\bibfnamefont {S.~D.}\ \bibnamefont {Smart}}, \
  and\ \bibinfo {author} {\bibfnamefont {A.}~\bibnamefont {Alavi}},\ }\href
  {\doibase 10.1021/acs.jctc.5b01190} {\bibfield  {journal} {\bibinfo
  {journal} {J. Chem. Theory Comput.}\ }\textbf {\bibinfo {volume} {12}},\
  \bibinfo {pages} {1245} (\bibinfo {year} {2016})}\BibitemShut {NoStop}%
\bibitem [{\citenamefont {Fosso-Tande}\ \emph {et~al.}(2016)\citenamefont
  {Fosso-Tande}, \citenamefont {Nguyen}, \citenamefont {Gidofalvi},\ and\
  \citenamefont {A.~Eugene~DePrince}}]{JCTC.v2dm}%
  \BibitemOpen
  \bibfield  {author} {\bibinfo {author} {\bibfnamefont {J.}~\bibnamefont
  {Fosso-Tande}}, \bibinfo {author} {\bibfnamefont {T.-S.}\ \bibnamefont
  {Nguyen}}, \bibinfo {author} {\bibfnamefont {G.}~\bibnamefont {Gidofalvi}}, \
  and\ \bibinfo {author} {\bibfnamefont {I.}~\bibnamefont
  {A.~Eugene~DePrince}},\ }\href {\doibase 10.1021/acs.jctc.6b00190} {\bibfield
   {journal} {\bibinfo  {journal} {J. Chem. Theory Comput.}\ }\textbf {\bibinfo
  {volume} {in press}} (\bibinfo {year} {2016}),\
  10.1021/acs.jctc.6b00190}\BibitemShut {NoStop}%
\bibitem [{\citenamefont {Andersson}\ \emph {et~al.}(1990)\citenamefont
  {Andersson}, \citenamefont {Malmqvist}, \citenamefont {Roos}, \citenamefont
  {Sadlej},\ and\ \citenamefont {Wolinski}}]{JPC.94.5483}%
  \BibitemOpen
  \bibfield  {author} {\bibinfo {author} {\bibfnamefont {K.}~\bibnamefont
  {Andersson}}, \bibinfo {author} {\bibfnamefont {P.~A.}\ \bibnamefont
  {Malmqvist}}, \bibinfo {author} {\bibfnamefont {B.~O.}\ \bibnamefont {Roos}},
  \bibinfo {author} {\bibfnamefont {A.~J.}\ \bibnamefont {Sadlej}}, \ and\
  \bibinfo {author} {\bibfnamefont {K.}~\bibnamefont {Wolinski}},\ }\href
  {\doibase 10.1021/j100377a012} {\bibfield  {journal} {\bibinfo  {journal} {J.
  Phys. Chem.}\ }\textbf {\bibinfo {volume} {94}},\ \bibinfo {pages} {5483}
  (\bibinfo {year} {1990})}\BibitemShut {NoStop}%
\bibitem [{\citenamefont {Andersson}, \citenamefont {Malmqvist},\ and\
  \citenamefont {Roos}(1992)}]{JCP.96.1218}%
  \BibitemOpen
  \bibfield  {author} {\bibinfo {author} {\bibfnamefont {K.}~\bibnamefont
  {Andersson}}, \bibinfo {author} {\bibfnamefont {P.}~\bibnamefont
  {Malmqvist}}, \ and\ \bibinfo {author} {\bibfnamefont {B.~O.}\ \bibnamefont
  {Roos}},\ }\href {\doibase 10.1063/1.462209} {\bibfield  {journal} {\bibinfo
  {journal} {J. Chem. Phys.}\ }\textbf {\bibinfo {volume} {96}},\ \bibinfo
  {pages} {1218} (\bibinfo {year} {1992})}\BibitemShut {NoStop}%
\bibitem [{\citenamefont {Forsberg}\ and\ \citenamefont
  {Malmqvist}(1997)}]{CPL.274.196}%
  \BibitemOpen
  \bibfield  {author} {\bibinfo {author} {\bibfnamefont {N.}~\bibnamefont
  {Forsberg}}\ and\ \bibinfo {author} {\bibfnamefont {P.-A.}\ \bibnamefont
  {Malmqvist}},\ }\href {\doibase 10.1016/S0009-2614(97)00669-6} {\bibfield
  {journal} {\bibinfo  {journal} {Chem. Phys. Lett.}\ }\textbf {\bibinfo
  {volume} {274}},\ \bibinfo {pages} {196} (\bibinfo {year}
  {1997})}\BibitemShut {NoStop}%
\bibitem [{\citenamefont {Ghigo}, \citenamefont {Roos},\ and\ \citenamefont
  {Malmqvist}(2004)}]{CPL.396.142}%
  \BibitemOpen
  \bibfield  {author} {\bibinfo {author} {\bibfnamefont {G.}~\bibnamefont
  {Ghigo}}, \bibinfo {author} {\bibfnamefont {B.~O.}\ \bibnamefont {Roos}}, \
  and\ \bibinfo {author} {\bibfnamefont {P.-A.}\ \bibnamefont {Malmqvist}},\
  }\href {\doibase 10.1016/j.cplett.2004.08.032} {\bibfield  {journal}
  {\bibinfo  {journal} {Chem. Phys. Lett.}\ }\textbf {\bibinfo {volume}
  {396}},\ \bibinfo {pages} {142} (\bibinfo {year} {2004})}\BibitemShut
  {NoStop}%
\bibitem [{\citenamefont {Kurtz}, \citenamefont {Stewart},\ and\ \citenamefont
  {Dieter}(1990)}]{JCC.11.82}%
  \BibitemOpen
  \bibfield  {author} {\bibinfo {author} {\bibfnamefont {H.~A.}\ \bibnamefont
  {Kurtz}}, \bibinfo {author} {\bibfnamefont {J.~J.~P.}\ \bibnamefont
  {Stewart}}, \ and\ \bibinfo {author} {\bibfnamefont {K.~M.}\ \bibnamefont
  {Dieter}},\ }\href {\doibase 10.1002/jcc.540110110} {\bibfield  {journal}
  {\bibinfo  {journal} {J. Comput. Chem.}\ }\textbf {\bibinfo {volume} {11}},\
  \bibinfo {pages} {82} (\bibinfo {year} {1990})}\BibitemShut {NoStop}%
\bibitem [{\citenamefont {Turney}\ \emph {et~al.}(2012)\citenamefont {Turney},
  \citenamefont {Simmonett}, \citenamefont {Parrish}, \citenamefont
  {Hohenstein}, \citenamefont {Evangelista}, \citenamefont {Fermann},
  \citenamefont {Mintz}, \citenamefont {Burns}, \citenamefont {Wilke},
  \citenamefont {Abrams}, \citenamefont {Russ}, \citenamefont {Leininger},
  \citenamefont {Janssen}, \citenamefont {Seidl}, \citenamefont {Allen},
  \citenamefont {Schaefer}, \citenamefont {King}, \citenamefont {Valeev},
  \citenamefont {Sherrill},\ and\ \citenamefont {Crawford}}]{WIRCMS.2.556}%
  \BibitemOpen
  \bibfield  {author} {\bibinfo {author} {\bibfnamefont {J.~M.}\ \bibnamefont
  {Turney}}, \bibinfo {author} {\bibfnamefont {A.~C.}\ \bibnamefont
  {Simmonett}}, \bibinfo {author} {\bibfnamefont {R.~M.}\ \bibnamefont
  {Parrish}}, \bibinfo {author} {\bibfnamefont {E.~G.}\ \bibnamefont
  {Hohenstein}}, \bibinfo {author} {\bibfnamefont {F.~A.}\ \bibnamefont
  {Evangelista}}, \bibinfo {author} {\bibfnamefont {J.~T.}\ \bibnamefont
  {Fermann}}, \bibinfo {author} {\bibfnamefont {B.~J.}\ \bibnamefont {Mintz}},
  \bibinfo {author} {\bibfnamefont {L.~A.}\ \bibnamefont {Burns}}, \bibinfo
  {author} {\bibfnamefont {J.~J.}\ \bibnamefont {Wilke}}, \bibinfo {author}
  {\bibfnamefont {M.~L.}\ \bibnamefont {Abrams}}, \bibinfo {author}
  {\bibfnamefont {N.~J.}\ \bibnamefont {Russ}}, \bibinfo {author}
  {\bibfnamefont {M.~L.}\ \bibnamefont {Leininger}}, \bibinfo {author}
  {\bibfnamefont {C.~L.}\ \bibnamefont {Janssen}}, \bibinfo {author}
  {\bibfnamefont {E.~T.}\ \bibnamefont {Seidl}}, \bibinfo {author}
  {\bibfnamefont {W.~D.}\ \bibnamefont {Allen}}, \bibinfo {author}
  {\bibfnamefont {H.~F.}\ \bibnamefont {Schaefer}}, \bibinfo {author}
  {\bibfnamefont {R.~A.}\ \bibnamefont {King}}, \bibinfo {author}
  {\bibfnamefont {E.~F.}\ \bibnamefont {Valeev}}, \bibinfo {author}
  {\bibfnamefont {C.~D.}\ \bibnamefont {Sherrill}}, \ and\ \bibinfo {author}
  {\bibfnamefont {T.~D.}\ \bibnamefont {Crawford}},\ }\href {\doibase
  10.1002/wcms.93} {\bibfield  {journal} {\bibinfo  {journal} {WIREs
  Computational Molecular Science}\ }\textbf {\bibinfo {volume} {2}},\ \bibinfo
  {pages} {556} (\bibinfo {year} {2012})}\BibitemShut {NoStop}%
\bibitem [{\citenamefont {Sun}(2016)}]{pyscf_github}%
  \BibitemOpen
  \bibfield  {author} {\bibinfo {author} {\bibfnamefont {Q.}~\bibnamefont
  {Sun}},\ }\href@noop {} {} (\bibinfo {year} {2016}),\ \bibinfo {note}
  {\textsc{pyscf}: Python module for quantum chemistry,
  \url{https://github.com/sunqm/pyscf}}\BibitemShut {NoStop}%
\bibitem [{\citenamefont {Edmiston}\ and\ \citenamefont
  {Ruedenberg}(1963)}]{RevModPhys.35.457}%
  \BibitemOpen
  \bibfield  {author} {\bibinfo {author} {\bibfnamefont {C.}~\bibnamefont
  {Edmiston}}\ and\ \bibinfo {author} {\bibfnamefont {K.}~\bibnamefont
  {Ruedenberg}},\ }\href {\doibase 10.1103/RevModPhys.35.457} {\bibfield
  {journal} {\bibinfo  {journal} {Rev. Mod. Phys.}\ }\textbf {\bibinfo {volume}
  {35}},\ \bibinfo {pages} {457} (\bibinfo {year} {1963})}\BibitemShut
  {NoStop}%
\bibitem [{\citenamefont {Wouters}\ \emph {et~al.}(2015)\citenamefont
  {Wouters}, \citenamefont {Poelmans}, \citenamefont {{De Baerdemacker}},
  \citenamefont {Ayers},\ and\ \citenamefont {{Van Neck}}}]{CPC.191.235}%
  \BibitemOpen
  \bibfield  {author} {\bibinfo {author} {\bibfnamefont {S.}~\bibnamefont
  {Wouters}}, \bibinfo {author} {\bibfnamefont {W.}~\bibnamefont {Poelmans}},
  \bibinfo {author} {\bibfnamefont {S.}~\bibnamefont {{De Baerdemacker}}},
  \bibinfo {author} {\bibfnamefont {P.~W.}\ \bibnamefont {Ayers}}, \ and\
  \bibinfo {author} {\bibfnamefont {D.}~\bibnamefont {{Van Neck}}},\ }\href
  {\doibase 10.1016/j.cpc.2015.01.007} {\bibfield  {journal} {\bibinfo
  {journal} {Comput. Phys. Commun.}\ }\textbf {\bibinfo {volume} {191}},\
  \bibinfo {pages} {235} (\bibinfo {year} {2015})}\BibitemShut {NoStop}%
\bibitem [{\citenamefont {Barcza}\ \emph {et~al.}(2011)\citenamefont {Barcza},
  \citenamefont {Legeza}, \citenamefont {Marti},\ and\ \citenamefont
  {Reiher}}]{PhysRevA.83.012508}%
  \BibitemOpen
  \bibfield  {author} {\bibinfo {author} {\bibfnamefont {G.}~\bibnamefont
  {Barcza}}, \bibinfo {author} {\bibfnamefont {O.}~\bibnamefont {Legeza}},
  \bibinfo {author} {\bibfnamefont {K.~H.}\ \bibnamefont {Marti}}, \ and\
  \bibinfo {author} {\bibfnamefont {M.}~\bibnamefont {Reiher}},\ }\href
  {\doibase 10.1103/PhysRevA.83.012508} {\bibfield  {journal} {\bibinfo
  {journal} {Phys. Rev. A}\ }\textbf {\bibinfo {volume} {83}},\ \bibinfo
  {pages} {012508} (\bibinfo {year} {2011})}\BibitemShut {NoStop}%
\bibitem [{\citenamefont {Bredas}\ \emph {et~al.}(1994)\citenamefont {Bredas},
  \citenamefont {Adant}, \citenamefont {Tackx}, \citenamefont {Persoons},\ and\
  \citenamefont {Pierce}}]{CR.94.243}%
  \BibitemOpen
  \bibfield  {author} {\bibinfo {author} {\bibfnamefont {J.~L.}\ \bibnamefont
  {Bredas}}, \bibinfo {author} {\bibfnamefont {C.}~\bibnamefont {Adant}},
  \bibinfo {author} {\bibfnamefont {P.}~\bibnamefont {Tackx}}, \bibinfo
  {author} {\bibfnamefont {A.}~\bibnamefont {Persoons}}, \ and\ \bibinfo
  {author} {\bibfnamefont {B.~M.}\ \bibnamefont {Pierce}},\ }\href {\doibase
  10.1021/cr00025a008} {\bibfield  {journal} {\bibinfo  {journal} {Chem. Rev.}\
  }\textbf {\bibinfo {volume} {94}},\ \bibinfo {pages} {243} (\bibinfo {year}
  {1994})}\BibitemShut {NoStop}%
\bibitem [{\citenamefont {Craig}\ \emph {et~al.}(1993)\citenamefont {Craig},
  \citenamefont {Cohen}, \citenamefont {Schrock}, \citenamefont {Silbey},
  \citenamefont {Puccetti}, \citenamefont {Ledoux},\ and\ \citenamefont
  {Zyss}}]{JACS.115.860}%
  \BibitemOpen
  \bibfield  {author} {\bibinfo {author} {\bibfnamefont {G.~S.~W.}\
  \bibnamefont {Craig}}, \bibinfo {author} {\bibfnamefont {R.~E.}\ \bibnamefont
  {Cohen}}, \bibinfo {author} {\bibfnamefont {R.~R.}\ \bibnamefont {Schrock}},
  \bibinfo {author} {\bibfnamefont {R.~J.}\ \bibnamefont {Silbey}}, \bibinfo
  {author} {\bibfnamefont {G.}~\bibnamefont {Puccetti}}, \bibinfo {author}
  {\bibfnamefont {I.}~\bibnamefont {Ledoux}}, \ and\ \bibinfo {author}
  {\bibfnamefont {J.}~\bibnamefont {Zyss}},\ }\href {\doibase
  10.1021/ja00056a006} {\bibfield  {journal} {\bibinfo  {journal} {J. Am. Chem.
  Soc.}\ }\textbf {\bibinfo {volume} {115}},\ \bibinfo {pages} {860} (\bibinfo
  {year} {1993})}\BibitemShut {NoStop}%
\end{thebibliography}%

\end{document}